\documentclass{article}[11 pt]
\usepackage{authblk} 
\usepackage{graphicx} 
\usepackage{amsmath,amssymb}
\usepackage{natbib}
\usepackage{verbatim}
\usepackage{multirow}
\usepackage[capposition=top]{floatrow} 
\usepackage{setspace}
\usepackage{silence}
\usepackage{hyperref}
\usepackage{makecell}
\usepackage{float}
\newcommand{\single}{\baselineskip 15pt}
\DeclareMathOperator{\logit}{logit}

\title{Optimal two-phase sampling designs for generalized raking estimators with multiple parameters of interest}
\author[1,*]{Jasper~B. Yang}
\author[2]{Bryan~E. Shepherd}
\author[3]{Thomas Lumley}
\author[1,4]{Pamela~A. Shaw}

\affil[1]{Department of Biostatistics, University of Washington, Seattle, WA, USA}
\affil[2]{Department of Biostatistics, Vanderbilt University, Nashville, Tennessee, USA}
\affil[3]{Department of Statistics, University of Auckland, Auckland, New Zealand}
\affil[4]{Biostatistics Division, Kaiser Permanente Washington Health Research Institute, Seattle, WA, USA}

\affil[*]{Corresponding Author Email: jbyang@uw.edu}

\date{\today}
\usepackage[dvipsnames]{xcolor}
\begin{document}
\doublespacing
\maketitle
\newpage

\label{firstpage}

\begin{abstract}
Large observational datasets, including those derived from electronic health records, are a valuable resource for medical research but are often affected by missingness, measurement error, and misclassification. Two-phase sampling with generalized raking (GR) estimation is an efficient and robust approach to statistical inference in such settings. In this approach, variables that are unavailable or measured with error in a large phase 1 cohort are obtained with higher-quality measurements in a phase 2 subsample. Previous research has studied optimal phase 2 sampling designs for inverse probability weighted (IPW) estimators in non-adaptive, multi-parameter settings, and for GR estimators in single-parameter settings. In this work, we extend these results by deriving optimal adaptive, multiwave sampling designs for IPW and GR estimators when multiple parameters are of interest. We propose several practical allocation strategies and evaluate their performance through extensive simulations and a data example from the Vanderbilt Comprehensive Care Clinic HIV Study. Our results show that independently optimizing allocation for each parameter improves efficiency over traditional case-control sampling. We also derive an integer-valued, A-optimal allocation method that typically outperforms independent optimization. Notably, we find that optimal designs for GR can differ substantially from those for IPW, and that this distinction can meaningfully affect estimator efficiency in the multiple-parameter setting. These findings offer practical guidance for future two-phase studies involving incomplete or error-prone data.
\end{abstract}

\maketitle

\section{Introduction}
\label{s:intro}

Routinely collected observational data are an increasingly popular resource for biomedical research, largely due to the widespread adoption of electronic health records (EHRs) \citep{Lee2020}. This is a promising trend, as EHR databases hold clinically relevant information for large cohorts and are cost-effective compared to traditional study data. However, because these data were not originally collected for research purposes, they often lack variables of interest, and the variables that are available are susceptible to measurement error and misclassification \citep{Weiskopf2013, Bell2020}. Neglecting these issues can lead to misleading research findings, which may in turn negatively affect patient health outcomes \citep{Floyd2012, Shortreed2019}.

A common approach in these settings is to supplement the initial dataset with a validation study where error-free and otherwise expensive measures are collected for a subset of the initial cohort. This approach can be viewed through the lens of two-phase sampling, where the inexpensive data are collected in phase 1, and the validation subset is gathered in phase 2 \citep{Neyman1938, Giganti2020, Shepherd2023}. To estimate parameters efficiently under two-phase sampling, two key components must be addressed: estimation and design. Estimation approaches have been well-studied in this context and can generally be categorized as model-based or design-based. The latter, which include Horvitz--Thompson inverse probability weighted (IPW) estimators and the generally more efficient augmented inverse-probability weighted (AIPW) estimators, are often preferred due to their robust properties \citep{Lumley2017, Tao2017, Han2021b,williamson2024assessing}. The second key component, design, is the focus of this work. In particular, we focus on the optimal design of phase 2 with respect to design-based estimators.

 It is well known that stratified sampling yields  efficient design-based estimators when individuals differ across readily available variables \citep{Cochran1977, yang2025improving}. Under stratified sampling, \cite{Neyman1934} showed that the variance of a population total (or mean) is approximately minimized when stratum sample sizes are proportional to the product of a stratum's standard deviation and size, and \cite{Wright2017} later presented an exactly optimal algorithm. \cite{Breslow2009a} extended this idea to regression coefficients by harnessing their asymptotic equivalence to totals of influence functions, and \cite{Chen2022} further extended this work for generalized raking estimators. Although the true influence functions for the population cannot be obtained in practice, a benefit of two-phase sampling is that they can be estimated from phase 1 data. Noting that the accuracy of this estimation will depend on the (unknown) correlation between the available phase 1 variables and the expensive phase 2 variables, some have proposed breaking up phase 2 sampling into multiple waves so that data collected in prior waves can be used to inform the design of subsequent waves \citep{Mcisaac2015, Han2021a}. Studies applying these multiwave methods have already been conducted with promising results \citep{Shepherd2023}.

The aforementioned research on two-phase designs has largely focused on maximizing efficiency with respect to a single parameter, but there is an emerging need for further understanding of efficiency with respect to multiple parameters in a single study. Recently, \cite{Sauer2021} and \cite{rivera2022} proposed methods for optimizing two-phase sampling for multiple parameters in cluster-correlated settings, focusing on minimizing the weighted sum of variances. These works acknowledge the challenge of estimating optimality parameters and incorporate pilot phase 2 samples in their examples; however, they do not fully investigate multiwave approaches, which allow for better estimation of design nuisance parameters by re-estimating them as validation data accumulate. Additionally, their methods focused on optimal designs for the Horvitz--Thompson IPW estimator, and their performance has not been evaluated alongside more efficient approaches such as generalized raking and its close relative, augmented inverse probability weighting (AIPW). In a related work, \cite{Wang2023} introduced a model-free approach for approximating optimal designs for multiple parameters under a maximin criterion. However, they also did not consider multiwave sampling, and their focus on Poisson sampling contrasts with the stratified sampling designs considered here.

In practice, multiwave sampling with generalized raking has been applied to multi-parameter problems, but often in ad hoc ways. For example, \cite{Shepherd2023} examined the association between maternal weight change during pregnancy and two outcomes, childhood obesity and asthma, using a two-phase, multiwave design. Their sampling was optimized separately by outcome, first allocating 75\% of the phase 2 sample for estimation of the primary obesity parameter and the remaining samples for the secondary asthma parameter. It is unlikely that this strategy was truly optimal for the research goals. There is large potential for other scenarios where researchers may want to answer multiple questions in one study, so understanding optimal design for these cases is needed. 

In this paper, we propose practical strategies for optimal stratified sampling for two-phase, multiwave studies using design-based estimators for multiple parameters of interest. These strategies include a simultaneous approach where optimum allocation is computed at each wave for every parameter of interest, a sequential approach where optimum allocation is computed at each wave for one parameter of interest, and an A-optimality approach where optimum allocation is computed at each wave to minimize the sum of variances for the multiple parameter estimates of interest. We also derive a novel extension of Wright's exact integer-valued allocation for one parameter to A-optimal allocation for multiple parameters. The rest of the paper is organized as follows. In Section 2 we review optimal design results when one parameter is of interest. In Section 3, we extend these results to the multiple parameter case and develop the three proposed strategies. These strategies are compared in a simulation study in Section 4 and are demonstrated in an example from the Vanderbilt Comprehensive Care Clinic HIV Study in Section 5. We conclude with summaries of our findings and recommendations for future studies.

\section{Optimal two-phase study designs for one parameter}
\label{s:review}

\subsection{Design-based estimators}
\label{s:neyman}
Design-based estimation is a robust approach to population inference that focuses on randomness due to the sampling process rather than the data-generating mechanism. A canonical example is the  Horvitz--Thompson inverse probability weighted (IPW) estimator for a population total $T_y = \sum_{i=1}^N y_i$ using a sample of size $n$. The IPW estimator in this case is $\hat{T}_{IPW}  = \sum_{i=1}^N  R_i \pi^{-1}_i y_i$, where $R_i$ is the indicator for whether unit $i$ is sampled (so that $\sum_{i=1}^NR_i = n$) and the weight $\pi_i^{-1} := w_i$ is the inverse of $\pi_i$, the sampling probability of unit $i$. This estimator is unbiased for $T_y$ provided that the $\pi_i$'s are non-zero, and it requires no distributional assumptions on $Y$. 

In practice, the collection of $y_i$ on a sample of size $n$ is often supplemented with less expensive variables $y_i^*$ which are correlated with $y$ and available for the entire population of size $N$. In such cases, another design-based estimation approach is Generalized Raking (GR), which uses the known population totals of $y_i^*$ to improve IPW estimators for the total of $y_i$ \citep{Deville1993}. This improvement is achieved by calibrating sampling weights so that the weighted estimate of the population total for the auxiliary variable using the subsample matches the true population total. More formally, GR estimators adjust IPW weights by a factor of $g_i$ so that they become $w^*_i = g_i\pi_i^{-1}$ for individual $i$, and the estimator becomes $\hat{T}_\mathrm{GR} = \sum_{i=1}^N R_i w^*_i y_i, 
$
where the $g_i$ values are chosen such that
\begin{equation}\label{Calibrationeq}
    \sum_{i=1}^NR_iw^*_iy^*_i = \sum_{i=1}^Ny^*_i.
\end{equation}
\noindent Since there may be multiple values of $g_i$ which satisfy Equation (\ref{Calibrationeq}), the specific $g_i$ values are selected as those that minimize $\sum_{i=1}^NR_id(g_i/\pi_i, 1/\pi_i)$ for distance measure $d(a,b) = a\mathrm{log}(a/b) - a + b$. GR estimators are at least as asymptotically efficient as traditional IPW estimators and are more efficient when the calibration (or raking) variable $y^*$ is correlated with the variable of interest \citep{Breslow2009a}. Moreover, GR estimators are consistent under the same assumptions as IPW estimators and straightforward to implement in standard statistical software, so these efficiency gains come at little additional cost \cite{SurveyPackage}.

Compared to a population total, a more general target of estimation is an arbitrary scalar parameter $\beta$. In the two-phase framework, the goal of phase 2 is to efficiently estimate $\tilde{\beta}$, the estimator for $\beta$ that would have been obtained if all data were available for the phase 1 cohort. In this work, we assume that such a $\tilde{\beta}$ exists and is asymptotically linear, meaning it admits the form
\begin{equation}\label{eq:AsympLin}
\sqrt{N}(\tilde{\beta} - \beta) = \frac{1}{\sqrt{N}}\sum_{i = 1}^N h_i + o_p(1),
\end{equation}
where $h_i$ is the influence function for unit $i$. A common example is when $\beta$ is a regression coefficient, in which case the full-data regression estimator $\tilde{\beta}$ is asymptotically linear under standard regularity conditions. Accordingly, $\tilde{\beta}$ can be expressed asymptotically as $\beta + \frac{1}{N}\sum_{i=1}^N h_i$, so $\tilde{\beta}$ is asymptotically equivalent to a sum of influence functions. An IPW phase 2 estimator $\hat{\beta}_\mathrm{IPW}$ can in turn be represented by
\begin{equation}\label{eq:IPWAsympLin}
\sqrt{N}(\hat{\beta}_\mathrm{IPW} - \beta) = \frac{1}{\sqrt{N}}\sum_{i = 1}^N R_i \pi_i^{-1} h_i + o_p(1),\end{equation}
under the asymptotic regime where $N \rightarrow \infty$, $n \rightarrow \infty$ and $n/N \rightarrow \gamma \in (0,1)$. 

The form of a GR estimator $\hat{\beta}_\mathrm{GR}$ follows similarly to equation (\ref{eq:IPWAsympLin}), with weights $\pi_i^{-1}$ replaced by calibrated weights satisfying Equation (\ref{Calibrationeq}). In this setting, we write $h_i^*$ for the raking variable available for the entire phase 1 cohort. \cite{Breslow2009a} showed that the optimal $h_i^*$ is the conditional expectation of $h_i$ given the observed phase 1 data. GR estimators calibrated in this way are asymptotically efficient among design-based estimators and asymptotically equivalent to the optimal augmented IPW (AIPW) estimator of \citet{robins1994estimation} \citep{Lumley2011}. In practice, the optimal $h_i^*$ is unknown but can be estimated through regression or imputation \cite{kulich2004improving, Breslow2009b, Oh2021b, Han2021b}.

\subsection{Optimal designs for IPW estimators: Neyman allocation}\label{s:IPWoptimaldesign}
 Under stratified sampling with simple random sampling within strata, the population is partitioned into $K$ strata of size $N_k$, and a sample of size $n_k$ is drawn uniformly without replacement from each stratum $k$. The asymptotic representation of the IPW estimator in Equation (\ref{eq:IPWAsympLin}) then becomes $\hat{\beta}_\mathrm{IPW} \approx \beta + \frac{1}{N}\sum_{k=1}^K \sum_{i \in I_k} R_i \frac{N_k}{n_k}h_i,$ where $I_k$ is the set of indices in stratum $k$ so that $\pi_i = n_k/N_k$ for $i \in I_k$. Applying the classic optimal allocation result of \cite{Neyman1934} in this setting shows that the variance of this estimator for a sample of size $n$ is minimized when 
\begin{equation}\label{eq:NeymanAll}
n_k = n\frac{N_k \sigma_k}{\sum_{k=1}^{K}N_k\sigma_k},
\end{equation}
where $\sigma_k$ is the stratum-specific population standard deviation of $h_i$ \citep{Chen2020, Han2021a}. Although $\sigma_k$ is typically unknown in practice, it can often be approximated using previously collected data. One drawback of Neyman allocation is that the resulting stratum sample sizes are not whole numbers, and they may not add up to the exact total sample size after rounding. In addition, the optimal stratum sizes $n_k$ are not constrained to be less than $N_k$. \cite{Wright2012, Wright2017} presented an exactly optimal solution to this problem by decomposing the variance of the final stratified estimator into a difference of discrete terms. By selecting samples from strata corresponding to the largest of these terms, he derived an integer-valued algorithm which is hereafter referred to as Neyman--Wright allocation.
\subsection{Optimal designs for generalized raking}\label{S:OptimalDesignRaking}

\cite{Chen2022} derive the optimal design for GR estimators by noting their asymptotic equivalence to generalized regression (GREG) estimators. Let $T_h = \sum_{i=1}^N h_i$ be the total of influence functions among all $N$ units in phase 1. Then the GREG estimator for $T_h$ under stratified sampling is 
\[
    \hat{T}_{h, GREG} = \sum_{k=1}^K \sum_{i \in I_k} R_i \frac{N_k}{n_k}(h_i - \alpha - h_i^* \theta) + \sum_{i=1}^N (\alpha + h_i^* \theta), \nonumber
\]
where $\theta$ is the parameter from a least squares regression of $h$ on $h^*$ with an intercept $\alpha$. By the GR analog of Equation (\ref{eq:IPWAsympLin}), minimizing the variance of $\hat{T}_h$ is asymptotically equivalent to minimizing the variance of the estimator $\hat{\beta}$. If estimators for $\alpha$ and $\theta$ converge at a $\sqrt{n}$ rate, then the variance of the second summation is of smaller order than the variance of the first summation. It thus suffices to minimize the variance of the first summation. The first term is the Horvitz--Thompson IPW estimator of the population total of residuals from the least squares regression of $h$ on $h^*$, so, by the arguments of Section \ref{s:IPWoptimaldesign}, the optimal design for GR estimators is well-approximated by Neyman allocation for the total of these residuals.

Despite these differences in the optimal designs for raking and IPW estimators, \cite{Chen2022} found that they often lead to estimators with similar efficiency. Moreover, their optimal design for GR estimators involves both $h_i$ and $h_i^*$ which are not jointly available a priori. Consequently, their results suggest that optimizing sampling for an IPW estimator is reasonable even when using a GR estimator. However, we will later demonstrate that this reasoning does not necessarily extend to the multiple parameter setting.
\subsection{Multiwave sampling in two-phase designs}\label{s:multiwave}

Neyman allocation is particularly useful in two-phase settings where a strong relationship exists between the phase 1 data and validated data. In these cases, $\sigma_k$ in Equation (\ref{eq:NeymanAll}) can be closely approximated by fitting the target regression model on the phase 1 data to generate estimates $h_i^*$ of $h_i$ and using the within-stratum standard deviations of $h_i^*$ as estimates for $\sigma_k$ \citep{kulich2004improving, Breslow2009a}. Further, these $h_i^*$ can serve as informative auxiliary variables for GR estimation. In practice, however, phase 1 variables will not always be highly correlated with the true optimal allocation parameters. To handle such scenarios, \cite{Mcisaac2015} suggest dividing phase 2 into a series of subphases, which are often referred to as waves. Information about phase 2 variables collected in previous waves can then be used to inform sample allocation in subsequent waves. A further benefit of this adaptive, multiwave approach is that it allows strata to be subdivided during phase 2, thus offering another avenue for efficiency gain. Optimal stratification occurs when optimum allocation assigns an equal number of samples to each stratum \citep{Sarndal2003, Amorim2021}. Theoretically, one wave per sampling unit would ensure the maximum amount of information is available at each decision point, but this is rarely practical. Instead, a more modest choice of 2-6 waves has been found to be useful \citep{Shepherd2023}. The simulations in this work are carried out over 4 waves.

\section{Considerations for Multiple Parameters}
\label{s:MultipleOutcomes}

\subsection{Setup}
\label{s:setup}
We now consider the case where the parameter of interest is a vector $\boldsymbol{\beta} = (\beta_1, ..., \beta_P)^T$. Following a standard two-phase sampling setting, we assume that inexpensive variables have been collected for a phase 1 cohort of size $N$, but identification of $\boldsymbol{\beta}$ requires expensive variables that can only be collected for a fixed $n < N$ units in phase 2. We again assume that an asymptotically linear estimator $\tilde{\boldsymbol{\beta}}$ for $\boldsymbol{\beta}$ would be available if the phase 2 variables were observed for the entire cohort, as is the case in standard regression settings. Our aim is to determine a stratified sampling design for selecting the $n$ phase 2 samples that will lead to an efficient design-based estimator $\hat{\boldsymbol{\beta}}$ for $\boldsymbol{\beta}$.

Write $\mathbf{h}_i$ for the $P$-dimensional influence function vector corresponding to the estimator $\tilde{\boldsymbol{\beta}}$ of $\boldsymbol{\beta}$ for person $i$. By Equation (\ref{eq:AsympLin}), an estimator  for $\boldsymbol{\beta}$ if the phase 2 variables were observed for the entire phase 1 cohort can for large $N$ be well approximated by $\tilde{\boldsymbol\beta} = \boldsymbol{\beta} + \frac{1}{N}\sum_{i = 1}^N \mathbf{h}_i$. However, these variables, and hence this total, are unknown. Instead, we estimate the sum by conducting stratified random sampling to select $n_k$ samples from the $N_k$ in each stratum.

For stratum $k$, we define the within-stratum covariance matrix of $\mathbf{h}_i$ as $V_k$, a $P \times P$ matrix with $(j,l)$-th entry $\sigma_{jl,k} \equiv \mathrm{cov}_k(\mathbf{h}_j,\mathbf{h}_l)$. When $j=l$, we write $\sigma_{jl,k} = \sigma^2_{j,k} = \mathrm{var} (\mathbf{h}_{j})$. Letting $w_k = \frac{N_k}{n_k}$, we have the IPW estimator $\boldsymbol{\hat{\beta}}_\mathrm{IPW} = \tilde{\boldsymbol{\beta}} + \sum_k \sum_{i \in I_k} R_i w_k \mathbf{h}_i$ and
\begin{equation}\label{eq:IPWVarMult}
 \mathrm{var}(\boldsymbol{\hat{\beta}}_\mathrm{IPW}) = \sum_{k=1}^K \frac{N_k^2}{n_k} V_k - \sum_{k=1}^K N_k V_k.
 \end{equation}
 The second term, which reflects the finite population correction, only depends on the fixed $N_k$ and $V_k$, so the aim is to select the $n_k$'s that minimize the first term subject to $\sum_{k=1}^K n_k = n$.

\subsection{Simple strategies: independent allocation for each parameter}\label{s:simplestrat}
A natural approach to optimize allocation for multiple parameters in a single study is to divide the budget of $n$ phase 2 sampling units into $P$ pieces, one for each parameter of interest, and then determine the allocation for each group independently using known Neyman or Neyman--Wright allocation results for the IPW estimator. Let $n_p$ be the number of phase 2 samples to be allocated optimally with respect to the $p$-th parameter of interest, where $\sum_{p=1}^P n_p = n$. Also, let $n_{pk}$ denote the sample size for stratum $k$ with $\sum_{k=1}^K n_{pk} = n_p$. The overall optimal allocation for $n_k$ becomes
\begin{equation}\label{eq:3.2.1}
n_k = \sum_{p=1}^P n_{pk} = \sum_{p=1}^P n_p\frac{N_k \sigma_{p,k}}{\sum_{k=1}^{K}N_k\sigma_{p,k}}.
\end{equation}
This is equivalent to the sum of the individual Neyman allocations for each parameter of interest. Alternatively, one could derive and sum across the \cite{Wright2017} integer-valued optimal design. 

If phase 2 is conducted over multiple waves, the above allocation may be implemented in multiple ways. Let $n_{k(t)}$ be the sample size in stratum $k$ at wave $t = 1, ..., T$. Two possible approaches are:
\begin{itemize}
    \item Simultaneous approach: At each wave, re-calculate $n_{k(t)}$ according to Equation (\ref{eq:3.2.1}), or the Neyman--Wright equivalent. Sampling at each wave is conducted with respect to all outcomes of interest, and allocation for both outcomes can make equal use of all information from prior waves.
    \item Sequential approach: At each wave, calculate optimal allocation with respect to a single outcome using Equation (\ref{eq:NeymanAll}) or the Neyman--Wright equivalent, only changing which outcome dictates allocation between waves. 
    This is the approach used by  \cite{Shepherd2023}. One benefit of this approach is that different strata can be defined for each outcome.
\end{itemize}

\begin{figure}
\centerline{\includegraphics[width=6.25in]{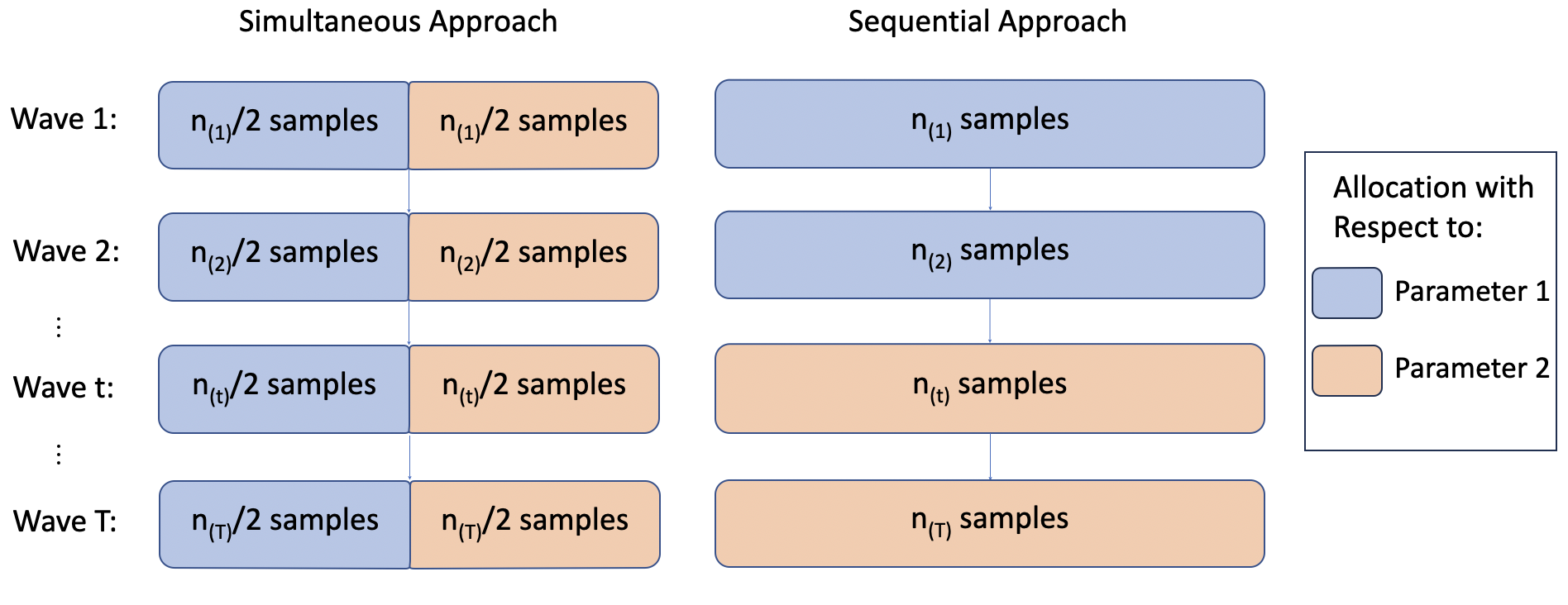}}
\caption{Visual depiction of simultaneous and sequential approaches when two outcomes are of equal interest.}
\label{fig:2}
\end{figure}

\subsection{Prioritized A-optimality}\label{s:weightedA}

The A-optimality approach is to minimize the sum of the variances for the $p$ parameter estimates \cite{Chan1982}. That is, to minimize $\mathrm{trace}\left(\mathrm{Var}(\boldsymbol{\hat{\beta}}) \right) = \sum_{p=1}^P\sum_{k=1}^K \frac{N^2_k}{n_k}\sigma^2_{p,k}$ in the IPW case. Because the parameters may have varying levels of importance towards the research goals, it may be useful to allow this sum to be weighted. Denote $a_p$ as the pre-defined priority weight for the $p$th parameter estimate, where $\sum_{p=1}^P a_p = 1$ for convenience. Traditional (non-prioritized) A-optimality now becomes a special case where all weights are equal. We want to find
\begin{equation}\label{eq:3.3.1}
\mathop{\mathrm{argmin}}_{n_1, \ldots, n_K} \sum_{p=1}^P\left[a_p\sum_{k=1}^K \frac{N^2_k}{n_k}\sigma^2_{p,k}\right] = \mathop{\mathrm{argmin}}_{n_1, \ldots, n_K} \sum_{k=1}^K\frac{N_k^2\sum_{p=1}^P a_p \sigma_{p,k}^2}{n_k},
\end{equation}
subject to the constraint $\sum_{k=1}^K n_k = n$. The continuous solution can be obtained using Lagrange multipliers (see \cite{rivera2022}) as
\begin{equation}
    n_k = n\frac{N_k\sqrt{\sum_{p=1}^P a_p \sigma^2_{p,k}}}{\sum_{k=1}^KN_k\sqrt{\sum_{p=1}^P a_p \sigma^2_{p,k}}}, \nonumber
\end{equation}
which is the Neyman allocation with the standard deviation of a single variable replaced by $\sqrt{\sum_{p=1}^P a_p \sigma^2_{p,k}}$. 

We can also extend the Neyman--Wright algorithm to obtain an exact, integer-valued solution to Equation (\ref{eq:3.3.1}) satisfying $n_k \geq 2$ (for unbiased variance estimation) and $n_k < N_k$. This allocation, whose derivation is provided in Supplemental Section \ref{ss:Proof}, is given by forming the array of priority values

\begin{eqnarray}\label{eq:WrightArray}
\displaystyle \frac{N_1 \sqrt{\sum_{p=1}^P a_p \sigma_{p,1}^2}}{\sqrt{2\cdot3}} \kern 15pt & 
\displaystyle \frac{N_1 \sqrt{\sum_{p=1}^P a_p \sigma_{p,1}^2}}{\sqrt{3\cdot4}} \kern 15pt &  
\displaystyle \frac{N_1 \sqrt{\sum_{p=1}^P a_p \sigma_{p,1}^2}}{\sqrt{4\cdot5}} \kern 20pt \cdots \nonumber \\
& & \kern -65pt \vdots \nonumber \\
\displaystyle \frac{N_k \sqrt{\sum_{p=1}^P a_p \sigma_{p,k}^2}}{\sqrt{2\cdot3}} \kern 15pt & 
\displaystyle \frac{N_k \sqrt{\sum_{p=1}^P a_p \sigma_{p,k}^2}}{\sqrt{3\cdot4}} \kern 15pt &  
\displaystyle \frac{N_k \sqrt{\sum_{p=1}^P a_p \sigma_{p,k}^2}}{\sqrt{4\cdot5}} \kern 20pt \cdots \\
& & \kern -65pt \vdots \nonumber \\
\displaystyle \frac{N_K \sqrt{\sum_{p=1}^P a_p \sigma_{p,K}^2}}{\sqrt{2\cdot3}} \kern 15pt & 
\displaystyle \frac{N_K \sqrt{\sum_{p=1}^P a_p \sigma_{p,K}^2}}{\sqrt{3\cdot4}} \kern 15pt &  
\displaystyle \frac{N_K \sqrt{\sum_{p=1}^P a_p \sigma_{p,K}^2}}{\sqrt{4\cdot5}} \kern 15pt \cdots \nonumber
\end{eqnarray}
and taking the number of samples assigned to stratum $k$ as 2 plus the number of priority values in the $k$-th row that are among the largest $n-2K$ in (\ref{eq:WrightArray}). In the context of multiple parameters, where the number of strata is typically large in order to accommodate the different parameters and the size of some strata are hence small, this algorithm proves especially useful because, unlike Neyman allocation, it incorporates the constraint $n_k <N_k$.

\subsection{Optimal design for generalized raking with multiple parameters}\label{S:optimaldesignrakingmultipleparams}

The approaches detailed so far for IPW estimators extend naturally to GR estimators by the arguments of Section \ref{S:OptimalDesignRaking}. Let $\epsilon_{i,p}$ be the residual from the least squares regression of $\mathbf{h}_{i,p}$ on the full set of raking variables $\mathbf{h}^*_i$. Denote by $\sigma^{(\epsilon)}_{p,k}$ the standard deviation of $\epsilon_{i,p}$ within stratum $k$. Then, the independent allocation and prioritized A-optimality strategies for GR estimators are given by replacing $\sigma_{p,k}$ with $\sigma_{p,k}^{(\epsilon)}$ in Equations (\ref{eq:3.2.1}) and (\ref{eq:WrightArray}) respectively.

Although \cite{Chen2022} showed that the optimal designs for IPW and GR often lead to similar efficiencies for a single parameter, this formulation suggests that the multiple parameter setting is more complicated. In the single parameter setting, the IPW and GR designs only differ when residualizing the influence functions on the calibration variables changes the distribution of within-stratum variances across strata. In the multiple parameters setting, the designs will also differ when the correlation between the raking variables and the true influence functions differ across parameters. For instance, if the optimal designs for the $p$-th and $p'$-th parameters are different, and $\mathbf{h}^*_{i,p}$ is highly correlated with $\mathbf{h}_{i,p}$ but $\mathbf{h}^*_{i,p'}$ and $\mathbf{h}_{i,p'}$ are weakly correlated, then the A-optimal design will closely resemble the optimal design for $\hat{\boldsymbol{\beta}}_{p'}$ since raking alone will lead to small variance for $\hat{\boldsymbol{\beta}}_p$. In the measurement error setting, this occurs when the variables relating to one parameter are measured with high error, but for another parameter they are measured with low error. The data example in Section \ref{s:DataExample} highlights this phenomenon. 

\section{Simulation Study}
\label{s:SimStudy}

We conducted a series of simulations to compare the performance of different allocation strategies in the context of two-phase, multiwave sampling with multiple parameters of interest. We considered the problem of efficiently estimating logistic regression coefficients from the following models:
\begin{eqnarray}\label{eq:simmodels}
Pr\left(Y_{1i}|X_{1i}, X_{2i},Z_{i}, Y_{2i}\right) &=& \mathrm{expit}(\beta_{01} + \beta_{11}X_{1i} + \beta_{21}X_{2i} + \beta_{31}Z_{i} + \beta_{41}Y_{2i}) \nonumber \\
Pr\left(Y_{2i}|X_{1i}, X_{2i}, Z_{i}\right) &=& \mathrm{expit}(\beta_{02} + \beta_{12}X_{1i} + \beta_{22}X_{2i} + \beta_{32}Z_{i}).
\end{eqnarray}

Motivated by real-world examples, we focused on the measurement error setting, which can be viewed as a special case of missing data in which the error-free phase 2 variables are missing at phase 1, but error-prone phase 1 variables are correlated with the missing values. Within this framework, we considered scenarios corresponding to different parameters of interest and features of the data-generating process including correlation levels and degree of measurement error in phase 1 variables (Table \ref{TabSetup}). These scenarios fall into three broad categories: Two Outcomes of interest (Scenario 2O), Two Predictors of interest (Scenario 2P), and Two Outcomes and Two Predictors of interest (Scenario 2O2P).




\begin{table}[!ht]
\caption{Simulation scenarios.}
\centering
\begin{tabular}{|l|l|l|l|l|l|}
     \cline{1-6} \multicolumn{2}{|c|}{\textbf{Scenario}}
    & \multicolumn{1}{c|}{\textbf{\shortstack{\rule{0pt}{2.6ex}Coefficients \\ of Interest}}} & \textbf{Cor($Y_1, Y_2$)} & \textbf{Cor($X_1, X_2$)} & \textbf{Error level} \\ \hline
\multicolumn{1}{|l|}{}
    & A & $\beta_{11}$, $\beta_{12}$ & Low  & -     & Low  \\ 
\multicolumn{1}{|l|}{Two Outcomes,} 
    & B & $\beta_{11}$, $\beta_{12}$ & Low  & -     & High \\ 
\multicolumn{1}{|l|}{One Predictor (2O)} 
    & C & $\beta_{11}$, $\beta_{12}$ & High & -     & Low  \\ 
\multicolumn{1}{|l|}{} 
    & D & $\beta_{11}$, $\beta_{12}$ & High & -     & High \\ \hline

\multicolumn{1}{|l|}{} 
    & A & $\beta_{11}$, $\beta_{21}$ & -    & Low   & Low  \\ 
\multicolumn{1}{|l|}{One Outcome,} 
    & B & $\beta_{11}$, $\beta_{21}$ & -    & Low   & High \\ 
\multicolumn{1}{|l|}{Two Predictors (2P)} 
    & C & $\beta_{11}$, $\beta_{21}$ & -    & High  & Low  \\ 
\multicolumn{1}{|l|}{} 
    & D & $\beta_{11}$, $\beta_{21}$ & -    & High  & High \\ \hline

\multicolumn{1}{|l|}{Two Predictors,} 
    & A & $\beta_{11}$, $\beta_{21}, \beta_{12}, \beta_{22}$ & Moderate & Moderate & Low  \\ 
\multicolumn{1}{|l|}{Two Outcomes (2O2P)} 
    & B & $\beta_{11}$, $\beta_{21}, \beta_{12}, \beta_{22}$ & Moderate & Moderate & High \\ \hline
\end{tabular}
\label{TabSetup}
\end{table}

    

Our two motivating examples fall under Scenario 2O, but can easily be extended to the others. The first is the Vanderbilt Comprehensive Care Clinic (VCCC) HIV/AIDS cohort, where the relationship between CD4 cell count and two binary outcomes, AIDS-defining event and death, are of each interest \citep{melekhin2010postpartum, Giganti2020}. These two outcomes are highly correlated because AIDS may lead to death, and similar causes may also lead to both AIDS and death independently. A second motivating example is the PCORI Maternal Weight study, where the relationship between maternal weight change during pregnancy and two outcomes with low correlation, childhood obesity and asthma, were of interest \citep{Shepherd2023}. In both studies, phase 1 consisted of error-prone measurements obtained from electronic health records, while phase 2 involved chart review to obtain error-free measurements for the variables of interest.

\subsection{Setup}\label{sub:Setup}

At each simulation iteration, we generated a phase 1 dataset with N = 10,000 samples including two continuous variables, $X_1$ and $X_2$, one binary covariate $Z$, and two binary outcome variables $Y_1$ and $Y_2$ generated according to the logistic models in (\ref{eq:simmodels}). We also simulated error-prone or proxy versions of each variable, denoted $X^*_1, X^*_2, Z^*, Y^*_1, Y^*_2$, and assumed only these versions were collected in phase 1. The measurement errors were simulated with correlation structures to reflect realistic error structures. We considered varying degrees of measurement error for the outcome and exposure(s) of interest. Further details on the simulation setup are provided in Supplemental Section \ref{SSim}. 

The key distinguishing features between scenarios are the parameters of interest and the level of correlation between $Y_1$ and $Y_2$ or between $X_1$ and $X_2$. Scenarios 2O-A and 2O-B reflect the case of nearly independent outcomes, as in the PCORI maternal weight study, and Scenarios 2O-C and 2O-D reflect the case of correlated outcomes, as in the VCCC study, where one is a predictor for the other. Across scenarios, the prevalences of $Y_1$ and $Y_2$ were maintained at approximately 0.2 and 0.4 respectively, representing rare and common outcomes. Further, all continuous covariates were generated with variance 1 to minimize the contribution of scale. In practice, this would involve standardizing the covariates. Throughout this simulation study, we assume that the variance of the coefficients of interest are of equal importance, i.e. the weights $a_p$ in equation (\ref{eq:3.3.1}) are all equal. 

\subsection{Sampling designs}\label{sub:SampDesign}
For each of the ten combinations of correlation and measurement error levels, five phase 2 sampling design strategies were implemented assuming a phase 2 budget of $n = 1,000$ validation samples:
\begin{itemize}
    \item \textbf{Strategy 1:} Case-control sampling in a single wave.
    \item \textbf{Strategy 2:} Simultaneous independent allocation for each parameter in four equally sized waves. At each wave, $n_{(t)} = 250$ samples were split evenly among the parameters of interest, with each subset allocated optimally for a single parameter (e.g., in the two-parameter case, 125 samples were optimized independently for each parameter). 
    \item \textbf{Strategy 3:} Sequential independent allocation for each parameter over four equally sized waves. At each wave, $n_{(t)} = 250$ samples were allocated optimally for a single parameter, with each parameter receiving an equal number of dedicated waves.
     \item \textbf{Strategy 4:} Strategy 3, but with the order of parameter-specific optimization reversed.
    \item \textbf{Strategy 5:} Prioritized A-optimality over four equally sized waves. 
\end{itemize}
Further details on each strategy are provided in Supplemental Section \ref{s:SuppStrats}. We also consider a supplemental setting where $n = 500$.

Estimation of the final regression parameters of interest was performed using both generalized raking and IPW. For strategies 2-5, strata were formed based on combinations of $Y^*_{1}, Y^*_{2}$, $X_1^*$ and $X_2^*$. These stratum boundaries remained unchanged throughout phase 2 sampling for simplicity. Optimum allocation for wave 1 was calculated using Neyman allocation of the estimated influence functions $\mathbf{h}_i^*$ from fitting the target model on the error-prone phase 1 data for both IPW and GR estimators. For subsequent waves under the IPW estimator, we performed IPW-optimal Neyman allocation using influence functions from the target model fit to the previously collected phase 2 data with IPW estimators. For GR estimators, we used the GR-optimal approach detailed in Section \ref{S:optimaldesignrakingmultipleparams}, with the estimated influence functions from fitting the target model on the phase 1 data as the auxiliary raking variables (see Supplemental Section \ref{ss:rakingprocedure} for details). The R packages \textquotesingle survey\textquotesingle\ and \textquotesingle optimall\textquotesingle\ were used to implement the estimation and design procedures respectively \citep{SurveyPackage, yang2025optimum}.  Each sampling strategy was run 2,500 times on each dataset, and the observed variances for each estimate were reported.

\subsection{Simulation Results}\label{sub:SimResults}

The empirical variances of the regression coefficient estimates for each estimator for simulation Scenarios 2O, 2P, and 2O2P are presented in Tables~\ref{TabScenario2O}, \ref{TabScenario2P}, and Supplemental Table S3 respectively. Across all scenarios, GR estimators outperformed IPW estimators for each sampling strategy, with the largest efficiency gain from raking observed under the non-optimal case-control design. In general, the efficiency improvement from raking relative to IPW under case-control sampling was comparable in magnitude to that achieved by A-optimal sampling with IPW estimators relative to case-control sampling, and combining the two provided an additional but more modest gain. Supplemental tables S4, S5, S6 report empirical mean squared errors (MSEs), which followed similar patterns to the empirical variances, and Supplemental Tables S8 and S9 show that the empirical coverage probabilities of estimated 95\% confidence intervals are close to the nominal 0.95 level across strategies. The trends across allocation strategies were consistent for both IPW and GR estimators, and results in this section focus on the latter unless noted otherwise.  

Across all ten simulation scenarios, multiwave adaptive sampling with Neyman allocation of influence functions (Strategies 2-5) yielded more efficient regression coefficient estimates than case-control sampling (Strategy 1). Table \ref{TabERE} reports the empirical relative efficiency (ERE) of the GR estimators for each strategy, where ERE is defined as the ratio of the total empirical variance of the estimated regression coefficients under the A-optimal design to that under the given design. A-optimal designs (Strategy 5) achieved the best empirical relative efficiency in every scenario except for 2P-B, where it was narrowly beat out by the sequential approach (Strategy 3) with an ERE of 1.01. The simultaneous allocation strategy (Strategy 2) yielded variances closest to those of the A-optimal design, though it generally performed slightly worse.

In Scenario 2O, the sequential allocation strategies (Strategies 3 and 4) often minimized variance for the coefficient optimized in the final two waves at the expense of efficiency for the coefficient optimized in the first two waves. For instance, Strategy 3, which optimized for $\beta_{12}$ in the last two waves of phase 2, was the most efficient design for $\beta_{12}$ but the least efficient (excluding case-control) for $\beta_{11}$ in all four sub-scenarios (Table \ref{TabScenario2O}). Similarly, Strategy 4 was the least efficient approach for $\beta_{12}$ in all four sub-scenarios but the most efficient for $\beta_{11}$ in only two. In Scenario 2O2P, the sequential approach (Strategy 3) led to the lowest ERE at both error levels.

\begin{table}
\caption{\label{TabScenario2O} Empirical variance ($\times 10^3$) for each strategy under Scenario 2O: Two outcomes of interest.}
\centering
\begin{tabular}{|l|l|l|l|l|l|l|l|l|}
\hline
\textbf{\shortstack{Corr \\ level}} & 
\textbf{\shortstack{Error \\ level}} &  
\textbf{\shortstack{Estim-\\ator}} &  
 & 
\textbf{\shortstack{Strat 1: \\ Case-\\control}} & 
\textbf{\shortstack{Strat 2: \\ Simult-\\aneous}} & 
\textbf{\shortstack{Strat 3: \\ $\beta_{12}$ last}} & 
\textbf{\shortstack{Strat 4: \\ $\beta_{11}$ last}} & 
\textbf{\shortstack{Strat 5: \\ A-\\optimal}} \\ 
\hline
Low & Low & IPW & var($\hat{\beta}_{11}$) &   5.58 & 2.54 & 3.05 & 2.56 & 2.71 \\ 
 & (A) &  & var($\hat{\beta}_{12}$)  & 6.17 & 2.71 & 2.41 & 3.18 & 2.53 \\ \cline{3-9}
 &  & GR & var($\hat{\beta}_{11}$)   & 2.25 & 2.01 & 2.14 & 1.87 & 1.87 \\ 
 &  &  & var($\hat{\beta}_{12}$)   & 2.62 & 2.15 & 1.95 & 2.39 & 2.03 \\ \cline{2-9}
 & High & IPW  & var($\hat{\beta}_{11}$) &   5.75 & 4.97 & 5.52 & 4.52 & 4.82 \\ 
 & (B) &  & var($\hat{\beta}_{12}$) &   5.30 & 3.83 & 3.69 & 4.23 & 3.62 \\ \cline{3-9}
 &  & GR & var($\hat{\beta}_{11}$)   & 4.58 & 4.32 & 5.10 &  4.49 & 4.35 \\ 
 &  &  & var($\hat{\beta}_{12}$)   & 4.05 & 3.58 & 3.50 & 4.01 & 3.55 \\ \hline
High & Low & IPW & var($\hat{\beta}_{11}$) &7.44 & 4.25 & 4.46 & 4.06 & 4.11\\ 
 & (C)  &  & var($\hat{\beta}_{12}$)  & 7.47 & 3.36 & 3.30 & 4.28 & 3.42 \\ \cline{3-9}
 &  & GR & var($\hat{\beta}_{11}$)   & 3.56 & 2.94 & 3.22 & 2.90 & 2.87 \\ 
 &  &  & var($\hat{\beta}_{12}$)   & 3.35 & 2.67 & 2.44 & 2.96 & 2.53 \\ \cline{2-9}
 & High & IPW  & var($\hat{\beta}_{11}$) &   8.64 & 7.55 & 7.96 & 7.42 & 7.46 \\ 
 & (D) &  & var($\hat{\beta}_{12}$) &  6.75 & 4.98 & 4.54 & 5.41 & 4.86 \\ \cline{3-9}
 &  & GR & var($\hat{\beta}_{11}$)   & 7.12 & 6.91 & 7.42 & 6.79 & 6.79 \\ 
 &  &  & var($\hat{\beta}_{12}$)   & 5.04 & 4.58 & 4.35 & 5.37 & 4.48 \\ \hline
\end{tabular}
\end{table}

\begin{table}
\caption{\label{TabScenario2P} Empirical variance ($\times 10^3$) for each strategy under Scenario 2P: Two predictors of interest.}
\centering
\begin{tabular}{|l|l|l|l|l|l|l|l|l|l|}
\hline
\textbf{\shortstack{Corr \\ level}} & 
\textbf{\shortstack{Error \\ level}} &  
\textbf{\shortstack{Estim-\\ ator}} &  
 & 
\textbf{\shortstack{Strat 1:\\ Case-\\control}} & 
\textbf{\shortstack{Strat 2:\\ Simult-\\aneous}} & 
\textbf{\shortstack{Strat 3: \\ $\beta_{22}$ last}} & 
\textbf{\shortstack{Strat 4: \\ $\beta_{12}$ last}} & 
\textbf{\shortstack{Strat 5: \\ A-\\optimal}} \\ 
\hline
Low& Low & IPW & var($\hat{\beta}_{12}$) &   6.38 & 4.67 & 4.83 & 4.68 & 4.65 \\ 
 & (A) &  & var($\hat{\beta}_{22}$)  & 8.16 & 6.03 & 5.69 & 5.97 & 5.83 \\ \cline{3-9}
 &  & GR & var($\hat{\beta}_{12}$)   & 3.93 & 3.64 & 3.60 & 3.58 & 3.38 \\ 
 &  & & var($\hat{\beta}_{22}$)   & 5.81 & 5.09 & 5.16 & 5.15 & 5.05 \\ \cline{2-9}
 & High & IPW  & var($\hat{\beta}_{12}$) &   6.94 & 5.93 & 5.65 & 6.03 & 5.92 \\ 
 & (B) &  & var($\hat{\beta}_{22}$) &  8.43 & 6.88 & 6.81 & 6.90 & 7.24 \\ \cline{3-9}
 &  & GR & var($\hat{\beta}_{12}$)   & 5.57 & 5.35 & 5.28 & 5.33 & 5.21 \\ 
 &  & & var($\hat{\beta}_{22}$)   & 6.98 & 6.51 & 6.50 & 6.54 & 6.68 \\ \hline
High & Low & IPW & var($\hat{\beta}_{12}$) &   11.14 & 8.56 & 8.90 & 9.05 & 9.00 \\ 
 & (C) &  & var($\hat{\beta}_{22}$)  & 12.60 & 9.67 & 10.26 & 9.95 & 10.00 \\ \cline{3-9}
 &  & GR & var($\hat{\beta}_{12}$)   & 7.74 & 7.05 & 7.15 & 7.12 & 6.69\\ 
 &  & & var($\hat{\beta}_{22}$)   & 10.02 & 9.09 & 9.02 & 8.69 & 8.36 \\ \cline{2-9}
 & High & IPW  & var($\hat{\beta}_{12}$) &   11.49 & 10.42 & 10.88 & 10.91 & 10.64 \\ 
 & (D) &  & var($\hat{\beta}_{22}$) &  13.55 & 11.72 & 12.50 & 11.84 & 11.51 \\ \cline{3-9}
 &  & GR & var($\hat{\beta}_{12}$)   & 9.34 & 9.98 & 9.51 & 9.79 & 9.50 \\ 
 &  & & var($\hat{\beta}_{22}$)   & 11.51 & 11.13 & 11.26 & 11.00 & 10.68 \\ \hline
\end{tabular}
\end{table}

\renewcommand{\arraystretch}{1.2}
\begin{table}
\caption{\label{TabERE} Empirical relative efficiency for sum of variances for each strategy under GR. Scenario 2O: Two Outcomes, One Predictor of interest, Scenario 2P: One Outcome, Two Predictors of interest, Scenario 2O2P: Two Outcomes, Two Predictors of interest.}
\centering
\begin{tabular}{|l|l|l|l|l|l|l|l|l|l|}
\hline
\multicolumn{2}{|c|}{\textbf{Scenario}}  & 
\textbf{\shortstack{Corr \\ level}} &  
\textbf{\shortstack{Error\\ level}} &  
\textbf{\shortstack{Strat 1: \\ Case-\\control}} & 
\textbf{\shortstack{Strat 2: \\ Simult- \\ aneous}} & 
\textbf{\shortstack{Strat 3: \\ Sequential \\ order 1}} & 
\textbf{\shortstack{Strat 4: \\ Sequential \\ order 2}} & 
\textbf{\shortstack{Strat 5: \\ A-\\optimal}} \\ 
\hline
2O & A & Low & Low &   0.80 & 0.94 & 0.95 & 0.91 & 1.00 \\ 
 & B &   & High  & 0.92 & 1.00 & 0.92 & 0.93 & 1.00 \\ 
 & C &  High & Low   & 0.78 & 0.96 & 0.96 & 0.92 & 1.00 \\ 
 & D&  & High & 0.93 & 0.98 & 0.96 & 0.93 & 1.00 \\ \hline
 2P & A& Low & Low & 0.87 & 0.97 & 0.96 & 0.97 & 1.00 \\ 
 & B &  & High  & 0.95 & 1.00 & 1.01 & 1.00 & 1.00 \\ 
 & C& High & Low   & 0.85 & 0.93 & 0.93 & 0.95 & 1.00 \\ 
 & D& & High & 0.97 & 0.96 & 0.97 & 0.97 & 1.00 \\ \hline
 2O2P & A&  Moderate & Low &  0.82 & 0.97 & 0.94 & - & 1.00  \\ 
 & B& & High & 0.93 & 0.99 & 0.94 & - & 1.00  \\ \hline
\end{tabular}
\end{table}
\renewcommand{\arraystretch}{1}

Empirical variances increased with higher levels of measurement error across all sampling strategies. Additionally, the efficiency gains from GR compared to IPW diminished in high-error settings, where the phase 1 raking variables were less informative. Similarly, the efficiency advantage of optimal allocation over case-control sampling (Strategy 1) declined as measurement error increased. The largest empirical relative efficiency gain from A-optimal design in a high-error scenario was 0.08 in Scenario 2O-B, while the gains in low-error scenarios were consistently larger than 0.08 and as high as 0.22 in Scenario 2O-C (Table \ref{TabERE}). 

The relative efficiency gains from adaptive, multiwave sampling were fairly consistent across different levels of correlation between the outcomes or predictors of interest. However, case-control sampling was particularly inefficient relative to adaptive A-optimal allocation in scenarios with low error and high correlation (Table \ref{TabERE}). Under these conditions, optimal allocation is especially effective because low error leads to precise estimation of optimality parameters, while high correlation aligns optimality across parameters.
Results for the supplemental scenario with $n=500$ (Supplemental Tables S10–S12) were broadly consistent with the primary findings.

\section{Example: Vanderbilt Comprehensive Care Clinic Study}
\label{s:DataExample}

We also assessed performance of the various optimum allocation strategies using real EHR data collected on 1595 HIV-positive persons who received care at the Vanderbilt Comprehensive Care Clinic (VCCC) between 1999 and 2013. This dataset served as a motivating example for our work, as both death and AIDS-defining event (ADE) have been considered as outcomes of interest in prior studies, particularly as they relate to CD4 count at the time of antiretroviral therapy (ART) initiation  \citep{melekhin2010postpartum,Giganti2020}. These three variables, along with many others covering demographic information, medical history, and laboratory test results, were collected routinely during patient visits to the clinic and stored in the EHR. Given the error-prone nature of EHR-derived data, the research team also conducted validation by chart review for the entire cohort, which revealed many errors in EHR-derived variables relating to the ADE analysis. Notably, time to ADE was measured incorrectly in the EHR for 34.5\% of patients, and 10.8\% were found to have an incorrect ADE status (sensitivity: 0.87; specificity: 0.89; positive predictive value: 0.35). In contrast, variables relating to the death analysis were more accurate in the EHR. Time to death was initially measured with error in only 21.3\% of patients, and the indicator for death did not have any errors at all. Further details on this data set and discrepancies between routinely-collected and validated versions can be found in \cite{Giganti2020}.

In this example, we used the validated data on the whole cohort to assess the performance of optimum allocation strategies assuming the validation budget had been limited to 200 individuals. We defined the ``true" parameter values for $\beta_\mathrm{CD4, death}$ and $\beta_\mathrm{CD4, ADE}$ as the coefficients obtained by fitting the following Cox regression models on the entire validated data set: 
\begin{eqnarray}
h_{\mathrm{Death}}(t|\mathrm{CD4, Age}) &=& h_{0\mathrm{, Death}}(t) \cdot \exp(\beta_{\mathrm{CD4, Death}} \cdot \mathrm{CD4} + \beta_{\mathrm{Age, Death}} \cdot \mathrm{Age}) \nonumber \\
h_{\mathrm{ADE}}(t|\mathrm{CD4, Age}) &=& h_{0\mathrm{, ADE}}(t) \cdot \exp(\beta_{\mathrm{CD4, ADE}} \cdot \mathrm{CD4} + \beta_{\mathrm{Age, ADE}} \cdot \mathrm{Age}). \nonumber
\end{eqnarray}
Notably, this setup differs from the simulation study in the previous section where the parameters of interest were logistic regression coefficients from the superpopulation from which phase 1 was a simple random sample. The model difference highlights the flexibility of the proposed approach since estimators for coefficients in both logistic and Cox regression are asymptotically linear \citep{lin1989robust}. The difference in sampling framework isolates the phase 2 contribution to the variance, but this is the only component of the variance that can be controlled through the phase 2 sampling design, so the optimal allocation results apply as before. We performed phase 2 sampling across four waves of 50 samples using each optimum allocation strategy and compared the empirical variances observed over 1,000 sampling iterations. The procedure was implemented separately under optimal allocation for IPW and for raking following the procedures described in Section \ref{s:SimStudy} and Supplemental Section \ref{ss:rakingprocedure}. We display the results for both IPW and GR estimators under each sampling approach in Table \ref{TabDataExample}. 

\subsection{Data example: Results}

\begin{table}
\caption{\label{TabDataExample} Empirical variance ($\times 10^6$) of estimators (with type of design optimality) under each strategy in VCCC data example.}
\begin{tabular}{|l|l|l|l|l|l|l|l|l|}
\hline
\textbf{\shortstack{Estim- \\ator}} & \textbf{Design} &   & \textbf{\shortstack{Strat 1: \\  Case- \\control }} & \textbf{\shortstack{Strat 2: \\ Simult- \\ aneous}} & \textbf{\shortstack{Strat 3: \\ ADE \\ 1st
}} & \textbf{\shortstack{Strat 4: \\ Death \\ 1st}} & \textbf{\shortstack{Strat 5: \\ A-optimal}} \\ \hline
 IPW & IPW- & var($\hat{\beta}_{\mathrm{ADE}}$) & 3.08 &  2.16   & 2.53  & 2.21 & 1.97 \\  & optimal  &  var($\hat{\beta}_{\mathrm{Death}}$) & 0.87  & 0.64  & 0.64  & 0.79 & 0.80 \\ \cline{1-1} \cline{2-8}  GR &  IPW- & var($\hat{\beta}_{\mathrm{ADE}}$)  & 2.62 &  1.99  & 2.28  & 2.07 & 1.74 \\ & optimal &   var($\hat{\beta}_{\mathrm{Death}}$)  & 0.05  & 0.07 & 0.06  & 0.08 & 0.07 \\ \hline
 IPW & GR- & var($\hat{\beta}_{\mathrm{ADE}}$) & 3.08 &  1.99   & 2.17  & 1.73 & 1.61 \\    & optimal & var($\hat{\beta}_{\mathrm{Death}}$) & 0.87  & 0.80  & 0.78  & 0.79 & 1.34 \\ \cline{1-1} \cline{2-8}  GR & GR- & var($\hat{\beta}_{\mathrm{ADE}}$)  & 2.62 &  1.74   & 1.92  & 1.37 & 1.26 \\  & optimal   & var($\hat{\beta}_{\mathrm{Death}}$)  & 0.05  & 0.05 & 0.05  & 0.07 & 0.08 \\ \hline
\end{tabular}
\end{table}

The data example results further demonstrate the utility of adaptive, multiwave sampling with generalized raking on a real-world dataset. A-optimality successfully minimized the sum of variances among the presented strategies and demonstrated significant efficiency gains over case-control sampling for both IPW and GR estimators. 

Notably, across all sampling designs, GR led to an 88–95\% reduction in variance for $\hat{\beta}_\mathrm{death}$ but only a 12–22\% reduction for $\hat{\beta}_\mathrm{ADE}$ (Table \ref{TabDataExample}). This discrepancy is a direct consequence of the stark difference in measurement error levels in the phase 1 variables, which were used to derive influence functions serving as raking variables in the GR analysis. These influence functions were strongly correlated with the true influence functions for the death model but not for the ADE model, making raking substantially more effective for the former. Accordingly, Table \ref{TabDataExample} shows that using an IPW-optimal design between waves results in a notably higher sum of variances for raking estimators compared to the GR-optimal approach. The GR-based A-optimal design, constructed by regressing phase 2 influence function estimates on their phase 1 counterparts, recognized that raking alone would yield large gains for $\hat{\beta}_\mathrm{death}$ and thus prioritized minimizing variance for $\hat{\beta}_\mathrm{ADE}$ through the sampling design by mimicking the univariate IPW-optimal design for ADE (Supplemental Figure S1). This design was the only approach that effectively addressed the imbalance in efficiency gains across parameters. This type of trade-off does not arise in the single-parameter setting, where \cite{Chen2022} point out that IPW-optimal designs are often approximately optimal for GR as well. Thus, this example demonstrates the heightened importance of using an A-optimal design tailored to raking in the multiple parameter setting compared to the single parameter setting.

\section{Discussion}

This work presents a number of practical sampling designs for two-phase studies aimed at estimating multiple parameters with minimal variance. Through extensive numerical studies, we investigate the performance of these designs and demonstrate the strong benefits of adaptive, multiwave sampling using Neyman or Neyman--Wright allocation of influence functions and generalized raking estimators. These simulations, supported by theoretical results in Section \ref{s:MultipleOutcomes}, also illustrate the utility of a novel application of A-optimality to multiwave stratified sampling. A-optimality shows a promising ability to minimize the sum of variances for both IPW and generalized raking estimators across a wide range of measurement error and correlation scenarios. To support implementation of this approach, we have added an A-optimal allocation feature to the R package \textquotesingle optimall\textquotesingle, available on CRAN \citep{yang2025optimum}. While A-optimality has previously been considered in a non-adaptive setting for IPW estimators, our derivation of an A-optimal design specifically tailored to raking estimators and exploration of adaptive implementations are unique contributions of this work. Notably, we show that this approach can yield sampling strategies that offer substantial improvements over those optimized for IPW estimators, a result which differs from findings in the single-parameter setting. The data example in Section \ref{s:DataExample} is just one illustration of this approach's practical strengths, particularly its ability to target sampling towards estimators for which the phase 1 variables provide poor proxies for the variables of interest.

While A-optimality effectively minimizes the sum of variances, our results also suggest that simplified approaches, such as independently optimizing allocation for each parameter in each wave or optimizing for one parameter per wave sequentially, often yield very similar estimator variances. 
Among these approaches, the simultaneous strategy generally outperformed the sequential strategies. A-optimality's key advantage is its ability to assign explicit weights to different parameters, ensuring a more principled allocation. In contrast, independent allocation approaches (Strategies 3-4) rely on the choice of arbitrary sample split points. This limitation is highlighted in the data example: when one parameter is harder to measure, simply splitting the phase 2 sample in half does not lead to balanced efficiency gains. Similarly, in sequential strategies, later-allocated parameters benefit disproportionately from earlier sampling waves. In both cases, an optimal split point likely exists, but it cannot be known during the design stage. A-optimality naturally accounts for these complexities, making it the most practical and flexible strategy.

It should be noted that these simulations used only a simplified subset of the steps that could be taken to achieve an optimal sample design. Further improvements could be made by selecting stratum boundaries so that optimum allocation divides samples to strata evenly, even subdividing strata between waves as necessary \citep{Shepherd2023, Sarndal2003}. Another avenue for design refinement is incorporating informative priors to improve the wave 1 allocation rather than relying solely on error-prone phase 1 data \citep{Chen2020}. Multiple imputation could also be used to construct more efficient raking estimators, particularly in settings where the phase 1 variables are poor surrogates for the phase 2 variables of interest \citep{Han2021b}. 

The potential for future work on efficient sampling designs for multiple outcomes is vast and increasingly important given the increasing use of EHR in medical research \citep{Lee2020}. One natural extension of this work would be to extend the simulations to a broader range of outcome models and incorporate design enhancements such as optimal stratification. Under these conditions, the advantages of alternative optimality criteria that account for parameter covariances may become more apparent. Examples include D-optimality, E-optimality, and I-optimality, the latter focusing on minimizing prediction variance \citep{Wald1943, Pukelsheim2006, Goos2016}. Still, the practical advantages of A-optimal designs with multiwave adaptive sampling and generalized raking make it a compelling approach, and the findings of this work provide a foundation for future research on optimal design and estimation in studies using error-prone or incomplete data. 
\section*{Acknowledgments} This project was supported by the U.S. National Institutes of Health (NIH) grants R37-AI131771 and  P30-AI110527. 
\section*{Data Availability} Data from the example in this paper may be obtained by contacting the corresponding author and appropriate data use agreements. Data are not publicly available due to privacy restrictions. 
\def\maintex{}

\bibliographystyle{chicago}
\bibliography{references}

@article{robins1994estimation,
  title={Estimation of regression coefficients when some regressors are not always observed},
  author={Robins, James M and Rotnitzky, Andrea and Zhao, Lue Ping},
  journal={Journal of the American Statistical Association},
  volume={89},
  number={427},
  pages={846--866},
  year={1994},
  publisher={Taylor \& Francis}
}

@unpublished{yang2025improving,
  title={Improving optimal subsampling through stratification},
  author={Yang, Jasper B and Lumley, Thomas and Shepherd, Bryan E and Shaw, Pamela A},
  note={Preprint arXiv:2512.20837 available at https://arxiv.org/abs/2512.20837},
  year={2025}
}

@unpublished{williamson2024assessing,
  title={Assessing treatment effects in observational data with missing confounders: A comparative study of practical doubly-robust and traditional missing data methods},
  author={Williamson, Brian D and Krakauer, Chloe and Johnson, Eric and Gruber, Susan and Shepherd, Bryan E and van der Laan, Mark J and Lumley, Thomas and Lee, Hana and Munoz, Jose J Hernandez and Zhao, Fengyu and others},
  note={Preprint arXiv:2412.15012 available at https://arxiv.org/abs/2412.15012},
  year={2024}
}

@article{kulich2004improving,
  title={Improving the efficiency of relative-risk estimation in case-cohort studies},
  author={Kulich, Michal and Lin, DY},
  journal={Journal of the American Statistical Association},
  volume={99},
  number={467},
  pages={832--844},
  year={2004},
  publisher={Taylor \& Francis}
}

@article{Lee2020,
  author = {Lee, S. and Xu, Y. and D'Souza, A. G. and Martin, E. A. and Doktorchik, C. and Zhang, Z. and Quan, H.},
  year = {2020},
  title = {Unlocking the potential of electronic health records for health research},
  journal = {International Journal of Population Data Science},
  volume = {5},
  number = {1},
  pages = {610-617}
}

@article{Floyd2012,
  author = {Floyd, J. S. and Heckbert, S. R. and Weiss, N. S. and Carrell, D. S. and Psaty, B. M.},
  year = {2012},
  title = {Use of administrative data to estimate the incidence of statin-related rhabdomyolysis},
  journal = {Journal of the American Medical Association},
  volume = {307},
  number = {15},
  pages = {1580-1582},
}

@article{Giganti2020,
  author = {Giganti, M. J. and Shaw, P. A. and Chen, G. and Bebawy, S. S. and Turner, M. M. and Sterling, T. R. and Shepherd, B. E.},
  year = {2020},
  title = {Accounting for dependent errors in predictors and time-to-event outcomes using electronic health records, validation samples, and multiple imputation},
  journal = {The Annals of Applied Statistics},
  volume = {14},
  number = {2},
  pages = {1045},
}

@article{Shepherd2023,
  title={Multiwave validation sampling for error-prone electronic health records},
  author={Shepherd, Bryan E and Han, Kyunghee and Chen, Tong and Bian, Aihua and Pugh, Shannon and Duda, Stephany N and Lumley, Thomas and Heerman, William J and Shaw, Pamela A},
  journal={Biometrics},
  volume={79},
  number={3},
  pages={2649--2663},
  year={2023},
  publisher={Oxford University Press}
}

@article{melekhin2010postpartum,
  title={Postpartum discontinuation of antiretroviral therapy and risk of maternal {AIDS}-defining events, non-{AIDS}-defining events, and mortality among a cohort of {HIV}-1-infected women in the United States},
  author={Melekhin, Vlada V and Shepherd, Bryan E and Jenkins, Cathy A and Stinnette, Samuel E and Rebeiro, Peter F and Bebawy, Sally S and Rasbach, Daniel A and Hulgan, Todd and Sterling, Timothy R},
  journal={AIDS Patient Care and STDs},
  volume={24},
  number={5},
  pages={279--286},
  year={2010},
  publisher={Mary Ann Liebert, Inc. 140 Huguenot Street, 3rd Floor New Rochelle, NY 10801 USA}
}

@article{Lumley2011,
  author = {Lumley, T. and Shaw, P. A. and Dai, J. Y.},
  year = {2011},
  title = {Connections between survey calibration estimators and semiparametric models for incomplete data},
  journal = {International Statistical Review},
  volume = {79},
  number = {2},
  pages = {200-220},
}

@article{Chen2020,
  author = {Chen, T. and Lumley, T.},
  year = {2020},
  title = {Optimal multiwave sampling for regression modeling in two- phase designs},
  journal = {Statistics in Medicine},
  volume = {39},
  number = {30},
  pages = {4912-4921},
}

@book{Cochran1977,
  author = {Cochran, W. G.},
  year = {1977},
  title = {Sampling techniques},
  publisher = {John Wiley \& Sons},
}

@article{Neyman1934,
  author = {Neyman, J.},
  year = {1934},
  title = {On the Two Different Aspects of the Representative Method: The Method of Stratified Sampling and the Method of Purposive Selection},
  journal = {Journal of the Royal Statistical Society},
  volume = {97},
  number = {4},
  pages = {558-606},
}

@article{Breslow2009a,
  author = {Breslow, N. E. and Lumley, T. and Ballantyne, C. M. and Chambless, L. E. and Kulich, M.},
  year = {2009},
  title = {Improved {Horvitz--Thompson} estimation of model parameters from two-phase stratified samples: applications in epidemiology},
  journal = {Statistics in Biosciences},
  volume = {1},
  pages = {32-49},
}

@article{Mcisaac2015,
  author = {McIsaac, M. A. and Cook, R. J.},
  year = {2015},
  title = {Adaptive sampling in two-phase designs: a biomarker study for progression in arthritis},
  journal = {Statistics in Medicine},
  volume = {34},
  number = {21},
  pages = {2899-2912},
}

@article{Han2021a,
  author = {Han, K. and Lumley, T. and Shepherd, B. E. and Shaw, P. A.},
  year = {2021},
  title = {Two-phase analysis and study design for survival models with error-prone exposures},
  journal = {Statistical Methods in Medical Research},
  volume = {30},
  number = {3},
  pages = {857-874},
}

@article{Oh2021b,
  author = {Oh, E. J. and Shepherd, B. E. and Lumley, T. and Shaw, P. A.},
  year = {2021},
  title = {Raking and regression calibration: Methods to address bias from correlated covariate and time-to-event error},
  journal = {Statistics in Medicine},
  volume = {40},
  number = {3},
  pages = {631-649},
}

@article{Wright2012,
  author = {Wright, T.},
  year = {2012},
  title = {The equivalence of {Neyman} optimum allocation for sampling and equal proportions for apportioning the {US House of Representatives}},
  journal = {The American Statistician},
  volume = {66},
  number = {4},
  pages = {217-224},
}

@article{Wright2017,
  author = {Wright, T.},
  year = {2017},
  title = {Exact optimal sample allocation: More efficient than {Neyman}},
  journal = {Statistics \& Probability Letters},
  volume = {129},
  pages = {50-57},
}

@article{Neyman1938,
  author = {Neyman, J.},
  year = {1938},
  title = {Contribution to the theory of sampling human populations},
  journal = {Journal of the American Statistical Association},
  volume = {33},
  pages = {101-116},
}

@article{Amorim2021,
  author = {Amorim, G. and Tao, R. and Lotspeich, S. and Shaw, P. A. and Lumley, T. and Shepherd, B. E.},
  year = {2021},
  title = {Two-phase sampling designs for data validation in settings with covariate measurement error and continuous outcome},
  journal = {Journal of the Royal Statistical Society Series A: Statistics in Society},
  volume = {184},
  number = {4},
  pages = {1368-1389},
}

@book{Sarndal2003,
  author = {S\"arndal, C. E. and Swensson, B. and Wretman, J.},
  year = {2003},
  title = {Model assisted survey sampling},
  publisher = {Springer Science \& Business Media},
}

@unpublished{Lumley2017,
  title={Robustness of semiparametric efficiency in nearly-true models for two-phase samples},
  author={Lumley, Thomas},
  year={2017},
note={Pre-print arXiv:1707.05924, available at \url{https://arxiv.org/abs/1707.05924}}
}

@article{Tao2017,
  author = {Tao, R. and Zeng, D. and Lin, D. Y.},
  year = {2017},
  title = {Efficient semiparametric inference under two-phase sampling, with applications to genetic association studies},
  journal = {Journal of the American Statistical Association},
  volume = {112},
  number = {520},
  pages = {1468-1476},
}

@article{Han2021b,
  author = {Han, K. and Shaw, P. A. and Lumley, T.},
  year = {2021},
  title = {Combining multiple imputation with raking of weights: An efficient and robust approach in the setting of nearly true models},
  journal = {Statistics in Medicine},
  volume = {40},
  number = {30},
  pages = {6777-6791},
}

@article{Deville1993,
  author = {Deville, J. C. and Sarndal, C. E. and Sautory, O.},
  year = {1993},
  title = {Generalized raking procedures in survey sampling},
  journal = {Journal of the American Statistical Association},
  volume = {88},
  number = {423},
  pages = {1013-1020},
}

@article{Breslow2009b,
  author = {Breslow, N. E. and Lumley, T. and Ballantyne, C. M. and Chambless, L. E. and Kulich, M.},
  year = {2009},
  title = {Using the whole cohort in the analysis of case-cohort data},
  journal = {American Journal of Epidemiology},
  volume = {169},
  number = {11},
  pages = {1398-1405},
}

@Misc{SurveyPackage,
    author = {Thomas Lumley},
    year = {2024},
    title = {Survey: analysis of complex survey samples},
    note = {R package version 4.4},
    URL = {https://CRAN.R-project.org/package=survey}
  }

@article{Chen2022,
  author = {Chen, T. and Lumley, T.},
  year = {2022},
  title = {Optimal sampling for design-based estimators of regression models},
  journal = {Statistics in Medicine},
  volume = {41},
  number = {8},
  pages = {1482-1497},
}

@article{Chan1982,
  author = {Chan, N. N.},
  year = {1982},
  title = {A-optimality for regression designs},
  journal = {Journal of Mathematical Analysis and Applications},
  volume = {87},
  number = {1},
  pages = {45-50},
}

@article{yang2025optimum,
  title={Optimum Allocation for Adaptive Multi-Wave Sampling in {R}: The {R} Package optimall},
  author={Yang, Jasper B and Lumley, Thomas and Shepherd, Bryan E and Shaw, Pamela A},
  journal={Journal of Statistical Software},
  volume={114},
  pages={1--31},
  year={2025}
}

@article{Wald1943,
  author = {Wald, A.},
  year = {1943},
  title = {On the efficient design of statistical investigations},
  journal = {The Annals of Mathematical Statistics},
  volume = {14},
  number = {2},
  pages = {134-140},
}

@book{Pukelsheim2006,
  author = {Pukelsheim, F.},
  year = {2006},
  title = {Optimal design of experiments},
  publisher = {Society for Industrial and Applied Mathematics},
}

@article{Goos2016,
  author = {Goos, P. and Jones, B. and Syafitri, U.},
  year = {2016},
  title = {I-optimal design of mixture experiments},
  journal = {Journal of the American Statistical Association},
  volume = {111},
  number = {514},
  pages = {899-911},
}

@unpublished{Wang2023,
    title={A maximin optimal approach for model-free sampling designs in two-phase studies},
  author={Wang, Ruoyu and Wang, Qihua and Miao, Wang},
  note={Preprint arXiv:2312.10596 available at \url{https://arxiv.org/abs/2312.10596}},
  year={2023}
}

@article{Sauer2021,
  title={Optimal allocation in stratified cluster-based outcome-dependent sampling designs},
  author={Sauer, Sara and Hedt-Gauthier, Bethany and Haneuse, Sebastien},
  journal={Statistics in Medicine},
  volume={40},
  number={18},
  pages={4090--4107},
  year={2021},
  publisher={Wiley Online Library}
}

@article{rivera2022,
  title={Optimal sampling allocation for outcome-dependent designs in cluster-correlated data settings},
  author={Rivera-Rodriguez, Claudia and Haneuse, Sebastien and Sauer, Sara},
  journal={Statistical Methods in Medical research},
  volume={31},
  number={12},
  pages={2400--2414},
  year={2022},
  publisher={SAGE Publications Sage UK: London, England}
}

@article{Shortreed2019,
  title={Challenges and opportunities for using big health care data to advance medical science and public health},
  author={Shortreed, Susan M and Cook, Andrea J and Coley, R Yates and Bobb, Jennifer F and Nelson, Jennifer C},
  journal={American Journal of Epidemiology},
  volume={188},
  number={5},
  pages={851--861},
  year={2019},
  publisher={Oxford University Press}
}

@article{Bell2020,
  title={Frequency and types of patient-reported errors in electronic health record ambulatory care notes},
  author={Bell, Sigall K and Delbanco, Tom and Elmore, Joann G and Fitzgerald, Patricia S and Fossa, Alan and Harcourt, Kendall and Leveille, Suzanne G and Payne, Thomas H and Stametz, Rebecca A and Walker, Jan and others},
  journal={JAMA Network Open},
  volume={3},
  number={6},
  pages={e205867--e205867},
  year={2020},
  publisher={American Medical Association}
}

@article{Weiskopf2013,
  title={Methods and dimensions of electronic health record data quality assessment: enabling reuse for clinical research},
  author={Weiskopf, Nicole Gray and Weng, Chunhua},
  journal={Journal of the American Medical Informatics Association},
  volume={20},
  number={1},
  pages={144--151},
  year={2013},
  publisher={BMJ Group}
}

@article{lin1989robust,
  title={The robust inference for the Cox proportional hazards model},
  author={Lin, Danyu Y and Wei, Lee-Jen},
  journal={Journal of the American statistical Association},
  volume={84},
  number={408},
  pages={1074--1078},
  year={1989},
  publisher={Taylor \& Francis}
}
\newpage
\ifdefined\maintex
\renewcommand{\thesection}{S\arabic{section}}
\renewcommand{\thetable}{S\arabic{table}}  
\renewcommand{\thefigure}{S\arabic{figure}}
\renewcommand{\figurename}{Supplemental Figure} 
\renewcommand{\tablename}{Supplemental Table} 
\renewcommand{\theequation}{S\arabic{equation}}
\setcounter{table}{0}
\setcounter{figure}{0}
\setcounter{section}{0}
\setcounter{tocdepth}{2}
\else
\documentclass{article}
\usepackage{authblk}
\usepackage{multirow}
\usepackage{graphicx} 
\usepackage{subcaption} 

\usepackage{amsmath,amssymb}
\usepackage{natbib}
\usepackage{longtable}
\usepackage{makecell}
\usepackage{array}
\usepackage[margin=0.75in]{geometry}
\usepackage[capposition=top]{floatrow} 
\usepackage[colorlinks=true,linkcolor=cyan,urlcolor=cyan]{hyperref} 
\hypersetup{
    colorlinks,
    citecolor=black,
    filecolor=black,
    linkcolor=black,
    urlcolor=black
}
\usepackage{silence} 
\usepackage[dvipsnames]{xcolor}

\usepackage{tikz}
\usepackage{float}
\usetikzlibrary{shapes.geometric, arrows.meta, decorations,decorations.markings}
\tikzstyle{standard} = [rectangle, rounded corners, minimum width=2cm, minimum height=1cm,text centered, draw=black]
\tikzstyle{arrow} = [thick,-latex]

\WarningFilter*{latex}{Text page \thepage\space contains only floats}

\WarningFilter*{latex}{Float too large for page by}

\newcommand{\single}{\baselineskip 15pt}
\DeclareMathOperator{\logit}{logit}

\renewcommand{\thesection}{S\arabic{section}}
\renewcommand{\thetable}{S\arabic{table}}  
\renewcommand{\thefigure}{S\arabic{figure}}
\renewcommand{\figurename}{Supplemental Figure} 
\renewcommand{\tablename}{Supplemental Table} 
\renewcommand{\theequation}{S\arabic{equation}}
\setcounter{table}{0}
\setcounter{figure}{0}
\setcounter{tocdepth}{2}

\title{Supplementary Materials for ``Practical two-phase sampling designs for multiple parameters of interest''}

\author[1]{Jasper B. Yang}
\author[2]{Thomas Lumley}
\author[3]{Bryan~E. Shepherd}
\author[1,4]{Pamela A. Shaw}

\affil[1]{Department of Biostatistics, University of Washington, Seattle, WA, USA}
\affil[3]{Department of Biostatistics, Vanderbilt University, Nashville, Tennessee, USA}
\affil[2]{Department of Statistics, University of Auckland, Auckland, New Zealand}
\affil[4]{Biostatistics Division, Kaiser Permanente Washington Health Research Institute, Seattle, WA, USA}


\date{\today}

\begin{document}

\maketitle

\tableofcontents
\newpage

\fi

\section*{Supplemental Materials}

\section{Proof of exact integer-valued A-optimal allocation algorithm}\label{ss:Proof}

We seek an integer-valued allocation that satisfies
\[
\mathop{\mathrm{argmin}}_{n_1, \ldots, n_K} \sum_{p=1}^P\left[a_p\sum_{k=1}^K \frac{N^2_k}{n_k}\sigma^2_{p,k}\right] = \mathop{\mathrm{argmin}}_{n_1, \ldots, n_K} \sum_{k=1}^K\frac{N_k^2\sum_{p=1}^P a_p \sigma_{p,k}^2}{n_k},
\]
as well as the constraints $n_k \leq N_k$ and $n_k \geq 2$ for all $k = 1, ..., K.$ Note that $1-\frac{1}{n_k}$ can be re-written as a telescoping series $(1-\frac{1}{2})+(\frac{1}{2}-\frac{1}{3}) + \cdots + (\frac{1}{n_k - 1} - \frac{1}{n_k}) = \frac{1}{1 \cdot 2} + \frac{1}{2 \cdot 3} + \cdots + \frac{1}{(n_k - 1)\cdot n_k}$. Then,  

\begin{eqnarray}
\label{eq:S1}
\sum_{k=1}^K\frac{N_k^2\sum_{p=1}^P a_p \sigma_{p,k}^2}{n_k}
&=& \sum_{k=1}^K\left[\left(N_k^2\sum_{p=1}^P a_p \sigma_{p,k}^2\right)\left(1 - \frac{1}{1 \cdot 2} - \frac{1}{2 \cdot 3} - \cdots - \frac{1}{(n_k - 1)\cdot n_k}\right)\right] \nonumber \\
&=& \sum_{k=1}^K\left(N_k^2\sum_{p=1}^P a_p \sigma_{p,k}^2\right) \nonumber \\
&& -  \sum_{k=1}^K\left( \frac{N_k^2\sum_{p=1}^P a_p \sigma_{p,k}^2}{1\cdot2} + \cdots +  \frac{N_k^2\sum_{p=1}^P a_p \sigma_{p,k}^2}{(n_k-1)\cdot n_k}\right).
\end{eqnarray}
This expression, a weighted sum of variances, is minimized when the second sum in the difference is maximized. Thus, the optimum allocation occurs when the $n_k$ are chosen so that the $n$ largest terms are included in the second sum. For unbiased estimates of variance, we require that $n_k\geq 2$ for all $k$, so optimum allocation with this added constraint involves choosing the $n - 2K$ largest values from the array of priority values (note that the terms from Equation \ref{eq:S1} are square-rooted in (\ref{eq:S2}), which preserves the relative priorities but makes clear the similarity to Wright's original version):
\begin{eqnarray}\label{eq:S2}
\displaystyle \frac{N_1 \sqrt{\sum_{p=1}^P a_p \sigma_{p,1}^2}}{\sqrt{2\cdot3}} \kern 15pt & 
\displaystyle \frac{N_1 \sqrt{\sum_{p=1}^P a_p \sigma_{p,1}^2}}{\sqrt{3\cdot4}} \kern 15pt &  
\displaystyle \frac{N_1 \sqrt{\sum_{p=1}^P a_p \sigma_{p,1}^2}}{\sqrt{4\cdot5}} \kern 20pt \cdots \nonumber \\
& & \kern -65pt \vdots \nonumber \\
\displaystyle \frac{N_k \sqrt{\sum_{p=1}^P a_p \sigma_{p,k}^2}}{\sqrt{2\cdot3}} \kern 15pt & 
\displaystyle \frac{N_k \sqrt{\sum_{p=1}^P a_p \sigma_{p,k}^2}}{\sqrt{3\cdot4}} \kern 15pt &  
\displaystyle \frac{N_k \sqrt{\sum_{p=1}^P a_p \sigma_{p,k}^2}}{\sqrt{4\cdot5}} \kern 20pt \cdots \\
& & \kern -65pt \vdots \nonumber \\
\displaystyle \frac{N_K \sqrt{\sum_{p=1}^P a_p \sigma_{p,K}^2}}{\sqrt{2\cdot3}} \kern 15pt & 
\displaystyle \frac{N_K \sqrt{\sum_{p=1}^P a_p \sigma_{p,K}^2}}{\sqrt{3\cdot4}} \kern 15pt &  
\displaystyle \frac{N_K \sqrt{\sum_{p=1}^P a_p \sigma_{p,K}^2}}{\sqrt{4\cdot5}} \kern 15pt \cdots \nonumber.
\end{eqnarray}
The number of samples assigned to stratum $k$ is equal to 2 plus the number of priority values it has among the largest $n-2k$ in (\ref{eq:S2}). 

\section{Simulation Set-Up}\label{SSim}
\subsection{Data generation}
For every individual $i = 1, 2,\ldots, N$, covariates $X_1$, $X_2$, and $Z$ were simulated from a multivariate normal distribution with mean vector (0,0,0) and covariance matrix $$\left(\begin{matrix}1& \text{cor}(X_1, X_2) &0.10\\ \text{cor}(X_1, X_2)&1 & 0.25\\0.10&0.25&1\\\end{matrix}\right).$$Then, $Z$ was dichotomized into a binary variable and later standardized to have mean zero and variance 1. The two binary outcomes of interest, $Y_{1}$ and $Y_{2}$ were then generated according to the logistic models in Equation (\ref{eq:simmodels}). The coefficients varied by scenario and are presented in Supplemental Table \ref{tabS1}.
\subsection{Measurement error}
For each dataset, we assumed that error-prone or proxy versions of the variables $X_1, X_2, Z, Y_1$, and $ Y_2$, denoted $X^*_1, X^*_2, Z^*, Y^*_1,$ and $ Y^*_2$ respectively, were available in the phase 1 data. For the continuous variables $X_1$ and $X_2$, phase 1 versions were generated following the classical measurement error models $X_1^* = X_1 + U_{X_1}$ and $X_2^*=X_2+U_{X_2}$ respectively with $U_{X_1}$ and $U_{X_2}$ drawn from a multivariate normal distribution with mean zero. For binary variables $Z, Y_1,$ and $Y_2$, phase 1 versions were generated with pre-specified sensitivity and specificity. We simulated all of these errors with correlation to reflect the reality of error-prone data. We considered low and high degrees of measurement error for the outcome and exposure(s) of interest. The error-generating model parameters for $Y_1^*, Y_2^*$, $X_1^*$ and $X_2^*$ varied in each case, as shown in Supplemental Table \ref{tabS2}.

\subsection{Further details on design strategies 1-5} \label{s:SuppStrats}
The exact form of Strategies 1-5 varied depending on the simulation scenarion:
\begin{itemize}
\item Scenario 2O: Two outcomes, one covariate of interest
\begin{itemize}
    \item \textbf{Strategy 1:} Case-control sampling in a single wave. $n/4$ cases of $Y_1^*$ and $Y_2^*$, $n/4$ controls of $Y_1^*$ and cases of $Y_0^*$, $n/4$ cases of $Y_1^*$ and controls of $Y_0^*$, and $n/4$ controls of $Y_1^*$ and $Y_2^*$ were selected for sampling in a single wave.
    \item \textbf{Strategy 2:} Independent allocation for each parameter in four equally sized waves under the simultaneous approach. $n_{(t)}$ samples were collected at each wave, with $n_{(t)}/2$ allocated optimally with respect to $\beta_{11}$ and $n_{(t)}/2$ with respect to $\beta_{12}$.
    \item \textbf{Strategy 3:} Independent allocation for each parameter over four equally sized waves under the sequential approach. Waves 1 and 2 were allocated optimally with respect to $\beta_{11}$ and waves 3 and 4 with respect to $\beta_{12}$.
     \item \textbf{Strategy 4:} Same as Strategy 3, only Waves 1 and 2 were allocated optimally with respect to $\beta_{12}$ and waves 3 and 4 with respect to $\beta_{11}$.
    \item \textbf{Strategy 5:} Weighted A-optimality over four equally sized waves.
\end{itemize}
\item Scenario 2P: One outcome, two covariates of interest
\begin{itemize}
    \item \textbf{Strategy 1:} Case-control sampling in a single wave. $n/2$ cases of $Y_2^*$ and $n/2$ controls of $Y_2^*$ were selected for sampling in a single wave.
    \item \textbf{Strategy 2:} Independent allocation for each parameter in four equally sized waves under the simultaneous approach. $n_{(t)}$ samples were collected at each wave, with $n_{(t)}/2$ allocated optimally with respect to $\beta_{12}$ and $n_{(t)}/2$ with respect to $\beta_{22}$.
    \item \textbf{Strategy 3:} Independent allocation for each parameter over four equally sized waves under the sequential approach. Waves 1 and 2 were allocated optimally with respect to $\beta_{12}$ and waves 3 and 4 with respect to $\beta_{22}$.
     \item \textbf{Strategy 4:} Same as Strategy 3, only Waves 1 and 2 were allocated optimally with respect to $\beta_{22}$ and waves 3 and 4 with respect to $\beta_{12}$.
    \item \textbf{Strategy 5:} Weighted A-optimality over four equally sized waves.
\end{itemize}
\item Scenario 2O2P: Two outcomes, two covariates of interest
\begin{itemize}
    \item \textbf{Strategy 1:} Case-control sampling in a single wave. $n/4$ cases of $Y_1^*$ and $Y_2^*$, $n/4$ controls of $Y_1^*$ and cases of $Y_0^*$, $n/4$ cases of $Y_1^*$ and controls of $Y_0^*$, and $n/4$ controls of $Y_1^*$ and $Y_2^*$ were selected for sampling in a single wave.
    \item \textbf{Strategy 2:} Independent allocation for each parameter in four equally sized waves under the simultaneous approach. $n_{(t)}$ samples were collected at each wave, with $n_{(t)}/4$ allocated optimally with respect to $\beta_{11}$, $n_{(t)}/4$ with respect to $\beta_{12}$, $n_{(t)}/4$ with respect to $\beta_{12}$, and $n_{(t)}/4$ with respect to $\beta_{12}$.
    \item \textbf{Strategy 3:} Independent allocation for each parameter over four equally sized waves under the sequential approach. Wave 1 was allocated optimally with respect to $\beta_{11}$, wave 2 was allocated optimally with respect to $\beta_{21}$, wave 3 was allocated optimally with respect to $\beta_{12}$, and waves 4 with respect to $\beta_{22}$.
     \item \textbf{Strategy 4:} No strategy 4 in this case. Only one order was considered instead of all possible orderings of the four coefficients
    \item \textbf{Strategy 5:} Weighted A-optimality over four equally sized waves.
\end{itemize}
\end{itemize}

\subsection{Generalized raking procedure}\label{ss:rakingprocedure}
For the generalized raking procedures in the simulation study and data example, the raking variables $\mathbf{h}_i^*$ are the influence functions for the GR estimator of the target parameter that uses the error-prone phase 1 data. While theoretically, this may not be as efficient as using multiple imputation to approximate the optimal raking variable, it is more straightforward to implement, and several studies have found that it often led to similar performance in finite samples \citep{Oh2021b, Han2021b, Shepherd2023}. In computing the optimal raking design at each wave, we follow the procedures of Section \ref{S:optimaldesignrakingmultipleparams} with these raking variables as $\mathbf{h}_i^*$ and estimates for the true influence functions $\mathbf{h_i}$ computed as influence functions from the GR estimator using all phase 2 data that has been collected up to that point.

\section{Supplemental Figures}
\begin{figure}[!htb]
\centerline{\includegraphics[width=0.8\textwidth]{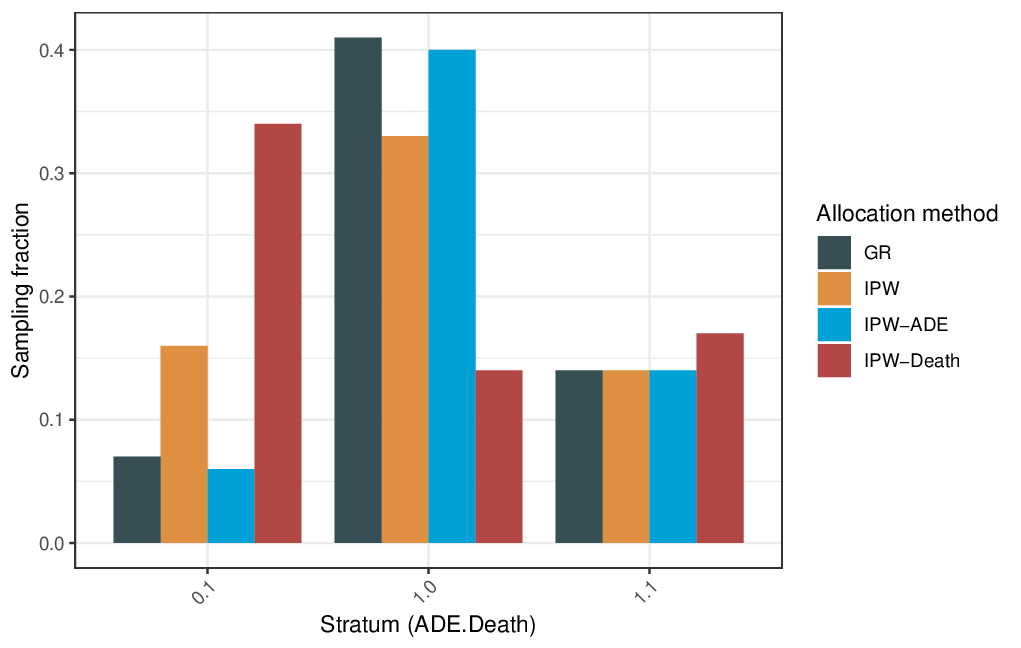}}
\caption{Optimal stratum fractions for three strata defined on ADE or Death cases under A-optimal allocation with respect to GR (GR), A-optimal allocation with respect to IPW (IPW), univariate IPW optimal allocation with respect to $\beta_\text{ADE}$ (IPW-ADE), and univariate IPW optimal allocation with respect to $\beta_{\text{death}}$ (IPW-Death).}
\label{fig:S1}
\end{figure}

\section{Supplemental Tables}

\begin{table}[!ht]
\centering
\caption {Coefficients for data-generating models in each simulation scenario.} 
\begin{tabular}{l|l|l|l|l|l|l|l|l|l|l|l|l|}
\cline{2-10}
    & \multicolumn{5}{c|}{Outcome 1} & \multicolumn{4}{c|}{Outcome 2} \\
    \cline{1-12}
    \multicolumn{1}{|l|}{Scenarios}& $\beta_{01}$ & $\beta_{11}$ & $\beta_{21}$ & $\beta_{31}$ & $\beta_{41}$ & $\beta_{02}$ & $\beta_{12}$ & $\beta_{22}$ & $\beta_{32}$ & cor($Y_1, Y_2$) & cor$(X_1,X_2)$  \\ \hline
\multicolumn{1}{|l|}{2O-A, 2O-B}     & -1.5           &   0.4           & 0             &  0.3  & 0          & -0.5         &    0.2          &    0.5          &     0         &   0.02    & -    \\ \hline
\multicolumn{1}{|l|}{2O-C, 2O-D}     & -3.1           &   0.4           & 1.0             &  0.7  & 1.9          & -0.8         &    0.2          &    1.3          &     0.8         &   0.44    & -       \\ \hline
\multicolumn{1}{|l|}{2P-A, 2P-B} & -     & -         & -          & -          & - & -2.1         & 0.3       & 0.7       & 0.7        & - & 0.05          \\ \hline
\multicolumn{1}{|l|}{ 2P-C, 2P-D} & -     & -         & -          & -          & - & -2.1         & 0.3       & 0.7       & 0.7         & - &  0.65        \\ \hline
\multicolumn{1}{|l|}{202P-A, 2O2P-B}          & -1.5         & 0.4       & 0.6       & 0.3         & 0.3 &-2.1     & 0.3         & 0.7          & 0.7          & 0.16 & 0.3          \\ \hline
\end{tabular}
\label{tabS1}
\end{table}

\begin{table}[ht]
\centering
\caption[Measurement error model parameters.]{Measurement error model parameters. Sens refers to sensitivity, the probability of the observed value being 1 if the true value is 1, and Spec to specificity, the probability of the true value being 0 if the true value is 0. The error model parameters for $Z$ were held constant across measurement error cases for simplicity, with $\text{Sens}(Z)$ = 0.9 and $\text{Spec}(Z)$ = 0.95.}
\begin{tabular}{l|l|l|l|l|l|l|l|}
    \cline{1-7}
    \multicolumn{1}{|l|}{\makecell{Scenarios}}  & Sens($Y_1$) & Spec($Y_1$) & Sens($Y_2$) & Spec($Y_2$) & Var($U_{X_1}$) & Var($U_{X_2}$)  \\ \hline
\multicolumn{1}{|l|}{2O-A, 2O-C}     & 0.95          &   0.99           & 0.9             &  0.95            & 0.15  & 0.1           \\ \hline
\multicolumn{1}{|l|}{2O-B, 2O-D}          & 0.85            &  0.90           &   0.8          &  0.85          & 0.5  & 0.1         \\ \hline
\multicolumn{1}{|l|}{2P-A, 2P-C}          & -            &  -           &   0.90         &  0.95          & 0.15  & 0.5          \\ \hline
\multicolumn{1}{|l|}{2P-B, 2P-D}          & -            &  -           &   0.85          &  0.90          & 0.4  & 0.6        \\ \hline
\multicolumn{1}{|l|}{2O2P-A}          & 0.95            &  0.99           &   0.90          &  0.95          & 0.15  & 0.5  \\ \hline
\multicolumn{1}{|l|}{2O2P-B}          & 0.85            &  0.90           &   0.80          &  0.85          & 0.4  & 0.6        \\ \hline
\end{tabular}
 \label{tabS2}
\end{table}

\newpage

\begin{table}[ht]
\caption{\label{TabScenario2O2P} Empirical variance ($\times 10^3$) under Scenario 2O2P: Multiple predictors and Multiple outcomes of interest.}
\centering
\begin{tabular}{|l|l|l|l|l|l|l|}
\hline 
\textbf{\shortstack{Error \\ level}} &  
\textbf{\shortstack{Estim-\\ator}} & &   
\textbf{\shortstack{Strat 1: \\ Case-\\control}} & 
\textbf{\shortstack{Strat 2: \\ Simult-\\aneous}} & 
\textbf{\shortstack{Strat 3: \\ Sequential}} & 
\textbf{\shortstack{Strat 5: \\ A-optimal}} \\ 
\hline
Low & IPW  & var($\hat{\beta}_{11}$) & 7.82 & 4.19 & 4.66 & 3.95  \\ 
 (A) &  & var($\hat{\beta}_{21}$)  & 8.84 & 5.48 & 6.19 & 5.40\\ 
 &  & var($\hat{\beta}_{12}$)   & 7.15 & 5.25 & 5.59 & 5.08 \\ 
 &  &  var($\hat{\beta}_{22}$)   & 8.85 & 6.53 & 7.24 & 6.54\\ \cline{2-7}
 & GR & var($\hat{\beta}_{11}$) &   2.93 & 2.47 & 2.82 & 2.45 \\ 
 &  &  var($\hat{\beta}_{21}$) &  5.25 & 4.02 & 4.67 & 4.08 \\ 
 &  & var($\hat{\beta}_{12}$)   & 4.68 & 4.01 & 3.76 & 3.65  \\ 
 &  & var($\hat{\beta}_{22}$)   & 6.37 & 5.73 & 5.62 & 5.66 \\ \hline
High & IPW  & var($\hat{\beta}_{11}$) &   8.03 & 5.76 & 6.60 & 6.16 \\ 
 (B) &  &   var($\hat{\beta}_{21}$)  & 9.19 & 6.82 & 7.53 & 6.91 \\ 
 &  & var($\hat{\beta}_{12}$) & 9.30 & 7.60 & 7.76 & 7.29  \\ 
 &  & var($\hat{\beta}_{22}$)   & 10.46 & 8.84 & 9.53 & 8.70  \\ \cline{2-7}
 & GR &var($\hat{\beta}_{11}$) &   5.73 & 5.01 & 5.91 & 5.31 \\ 
 &  &  var($\hat{\beta}_{21}$) &  7.29 & 6.14 & 6.91 & 6.24  \\ 
 &  & var($\hat{\beta}_{12}$)   & 6.86 & 7.02 & 6.88 & 6.73 \\ 
 &  &  var($\hat{\beta}_{22}$)   & 8.58 & 8.35 & 8.38 & 8.05 \\ \hline
\end{tabular}
\end{table}

\begin{table}[ht]
\caption{\label{TabMSEScenario2O} Empirical MSE ($\times 10^3$) under Scenario 2O: Multiple outcomes of interest.}
\centering
\begin{tabular}{|l|l|l|l|l|l|l|l|l|l|}
\hline
\textbf{\shortstack{Corr \\ level}} & 
\textbf{\shortstack{Error \\ level}} &  
\textbf{\shortstack{Estim-\\ator}} &  
 & 
\textbf{\shortstack{Strat 1: \\ Case-\\control}} & 
\textbf{\shortstack{Strat 2: \\ Simult-\\aneous}} & 
\textbf{\shortstack{Strat 3: \\ $\beta_{12}$ last}} & 
\textbf{\shortstack{Strat 4: \\ $\beta_{11}$ last}} & 
\textbf{\shortstack{Strat 5: \\ A-optimal}} \\ 
\hline
Low & Low & IPW & MSE($\hat{\beta}_{11}$) &   5.59 & 2.57 & 3.06 & 2.63 & 2.73 \\ 
 & (A)  &  & MSE($\hat{\beta}_{12}$)  & 6.18 & 2.74 & 2.42 & 3.19 & 2.53 \\ \cline{3-9}
 &  & GR & MSE($\hat{\beta}_{11}$)   & 2.25 & 2.02 & 2.14 & 1.89 & 1.88\\ 
 &  & & MSE($\hat{\beta}_{12}$)   & 2.62 & 2.17 & 1.96 & 2.41 & 2.04 \\ \cline{2-9}
 & High & IPW  & MSE($\hat{\beta}_{11}$) &   5.76 & 5.03 & 5.57 & 4.73 & 4.88 \\ 
 & (B) &  & MSE($\hat{\beta}_{12}$) &   5.30 & 3.88 & 3.72 & 4.25 & 3.63\\ \cline{3-9}
 &  & GR & MSE($\hat{\beta}_{11}$)   & 4.58 & 4.37 & 5.17 & 4.62 & 4.42 \\ 
 &  & & MSE($\hat{\beta}_{12}$)   & 4.05 & 3.63 & 3.54 & 4.05 & 3.58 \\ \hline
High & Low & IPW & MSE($\hat{\beta}_{11}$) &   7.44 & 4.30 & 4.47 & 4.09 & 4.11 \\ 
 & (C) &  & MSE($\hat{\beta}_{12}$)  & 7.48 & 3.36 & 3.30 & 4.28 & 3.42 \\ \cline{3-9}
 &  & GR & MSE($\hat{\beta}_{11}$)   & 3.56 & 2.95 & 3.22 & 2.95 & 2.88 \\ 
 &  & & MSE($\hat{\beta}_{12}$)   & 3.36 & 2.68 & 2.45 & 2.98 & 2.55 \\ \cline{2-9}
 & High & IPW  & MSE($\hat{\beta}_{11}$) &   8.64 & 7.70 & 7.98 & 7.69 & 7.60\\ 
 & (D) &  & MSE($\hat{\beta}_{12}$) &  6.76 & 4.99 & 4.56 & 5.47 & 4.90 \\ \cline{3-9}
 &  & GR & MSE($\hat{\beta}_{11}$)   & 7.12 & 6.98 & 7.45 & 7.06 & 6.92 \\ 
 &  & & MSE($\hat{\beta}_{12}$)   & 5.04 & 4.62 & 4.40 & 5.45 & 4.52 \\ \hline
\end{tabular}
\end{table}

\newpage

\begin{table}[ht]
\caption{\label{TabMSEScenario2P} Empirical MSE ($\times 10^3$) under Scenario 2P: Multiple predictors of interest.}
\centering
\begin{tabular}{|l|l|l|l|l|l|l|l|l|l|}
\hline
\textbf{\shortstack{Corr \\ level}} & 
\textbf{\shortstack{Error \\ level}} &  
\textbf{\shortstack{Estim-\\ator}} &  
 & 
\textbf{\shortstack{Strat 1: \\ Case-\\control}} & 
\textbf{\shortstack{Strat 2: \\ Simult-\\aneous}} & 
\textbf{\shortstack{Strat 3: \\ $\beta_{12}$ last}} & 
\textbf{\shortstack{Strat 4: \\ $\beta_{11}$ last}} & 
\textbf{\shortstack{Strat 5: \\ A-optimal}} \\ 
\hline
Low & Low & IPW & MSE($\hat{\beta}_{12}$) &   6.38 & 4.74 & 4.86 & 4.79 & 4.71\\ 
 & (A)  &  & MSE($\hat{\beta}_{22}$)  & 8.20 & 6.32 & 6.01 & 6.09 & 6.04 \\ \cline{3-9}
 &  & GR & MSE($\hat{\beta}_{12}$)   & 3.93 & 3.67 & 3.63 & 3.65 & 3.42 \\ 
 &  & & MSE($\hat{\beta}_{22}$)   & 5.84 & 5.33 & 5.66 & 5.30 & 5.32 \\ \cline{2-9}
 & High & IPW  & MSE($\hat{\beta}_{12}$) &   6.95 & 6.01 & 5.71 & 6.20 & 6.00 \\ 
 & (B) &  & MSE($\hat{\beta}_{22}$) &  8.45 & 7.38 & 7.68 & 7.09 & 7.76 \\ \cline{3-9}
 &  & GR & MSE($\hat{\beta}_{12}$)   & 5.57 & 5.47 & 5.36 & 5.51 & 5.30 \\ 
 &  & & MSE($\hat{\beta}_{22}$)   & 6.99 & 7.01 & 7.23 & 6.72 & 7.20 \\ \hline
High & Low & IPW & MSE($\hat{\beta}_{12}$) &   11.14 & 8.56 & 8.90 & 9.10 & 9.01 \\ 
 & (C) &  & MSE($\hat{\beta}_{22}$)  & 12.71 & 9.92 & 10.65 & 10.05 & 10.20\\ \cline{3-9}
 &  & GR & MSE($\hat{\beta}_{12}$)   & 7.74 & 7.06 & 7.14 & 7.14 & 6.69 \\ 
 &  & & MSE($\hat{\beta}_{22}$)   & 10.10 & 9.31 & 9.42 & 8.87 & 8.60 \\ \cline{2-9}
 & High & IPW  & MSE($\hat{\beta}_{12}$) &  11.51 & 10.46 & 10.88 & 11.00 & 10.67 \\ 
 & (D) &  & MSE($\hat{\beta}_{22}$) &  13.57 & 12.12 & 13.03 & 11.97 & 11.84\\ \cline{3-9}
 &  & GR & MSE($\hat{\beta}_{12}$)   & 9.34& 10.06 & 9.53 & 9.85 & 9.55 \\ 
 &  & & MSE($\hat{\beta}_{22}$)   & 11.53 & 11.38 & 11.76 & 11.16 & 10.96 \\ \hline
\end{tabular}
\end{table}
\newpage
\begin{table}[ht]
\caption{\label{TabMSEScenario2O2P} Empirical MSE ($\times 10^3$) under Scenario 2O2P: Multiple predictors and Multiple outcomes of interest.}
\centering
\begin{tabular}{|l|l|l|l|l|l|l|}
\hline
\textbf{\shortstack{Error \\ level}} &  
\textbf{\shortstack{Estim-\\ator}} &  
 & 
\textbf{\shortstack{Strat 1: \\ Case-\\control}} & 
\textbf{\shortstack{Strat 2: \\ Simult-\\aneous}} & 
\textbf{\shortstack{Strat 3: \\ Sequential}} & 
\textbf{\shortstack{Strat 5: \\ A-optimal}} \\ 
\hline
Low & IPW  & MSE($\hat{\beta}_{11}$) &   7.82 & 4.23 & 4.67 & 3.98  \\ 
 (A)&  & MSE($\hat{\beta}_{21}$)  & 8.89 & 5.69 & 6.38 & 5.61\\ 
 &  & MSE($\hat{\beta}_{12}$)   & 7.15 & 5.31 & 5.63 & 5.11  \\ 
 &  &  MSE($\hat{\beta}_{22}$)   & 8.89 & 6.87 & 8.21 & 7.05 \\ \cline{2-7}
 & GR & MSE($\hat{\beta}_{11}$) &   2.93 & 2.48 & 2.82 & 2.45 \\ 
 &  &  MSE($\hat{\beta}_{21}$) &  5.27 & 4.12 & 4.79 & 4.21 \\ 
  &  & MSE($\hat{\beta}_{12}$)   & 4.68 & 4.07 & 3.79 & 3.68  \\ 
 &  & MSE($\hat{\beta}_{22}$)   & 6.44 & 6.10 & 6.16 & 6.03 \\ \hline
High & IPW  & MSE($\hat{\beta}_{11}$) &  8.06 & 5.92 & 6.65 & 6.28 \\ 
(B) &  &   MSE($\hat{\beta}_{21}$)  & 9.21 & 7.05 & 7.80 & 7.14\\ 
 &  & MSE($\hat{\beta}_{12}$) & 9.30 & 7.78 & 8.12 & 7.44  \\ 
 &  & MSE($\hat{\beta}_{22}$)   & 10.52 & 9.52 & 11.39 & 9.43  \\ \cline{2-7}
 & GR &MSE($\hat{\beta}_{11}$) &   	
5.75 & 5.07 & 5.99 &  5.38 \\ 
 &  &  MSE($\hat{\beta}_{21}$) &  7.31 & 6.39 & 7.08 & 6.45 \\ 
 &  & MSE($\hat{\beta}_{12}$)   & 6.87 & 7.17 & 7.03 & 6.91\\ 
 &  &  MSE($\hat{\beta}_{22}$)   & 8.62 & 8.94 & 9.80 & 8.69 \\ \hline
\end{tabular}
\end{table}

\renewcommand{\arraystretch}{1.2}
\begin{table}[ht]
\caption{\label{TabS7} Empirical relative efficiency for sum of variances under IPW.}
\centering
\begin{tabular}{|l|l|l|l|l|l|l|l|l|l|}
\hline
\multicolumn{2}{|c|}{\textbf{Scenario}} &
\textbf{\shortstack{Corr \\ level}} &  
\textbf{\shortstack{Error \\ level}} &  
\textbf{\shortstack{Strat 1: \\ Case-\\control}} & 
\textbf{\shortstack{Strat 2: \\ Simult-\\aneous}} & 
\textbf{\shortstack{Strat 3: \\ Sequential \\ order 1}} & 
\textbf{\shortstack{Strat 4: \\ Sequential \\ order 2}} & 
\textbf{\shortstack{Strat 5: \\ A-optimal}} \\ 
\hline
2O & A & Low & Low &   0.45 & 1.00 & 0.96 & 0.91 & 1.00 \\ 
 & B & & High  & 0.76 & 0.96 & 0.92 & 0.96 & 1.00 \\ 
 & C & High & Low   & 0.50 & 0.99 & 0.97 & 0.90 & 1.00 \\ 
 & D & & High & 0.80 & 0.98 & 0.99 & 0.96 & 1.00 \\ \hline
 2P & A & Low & Low &   	
0.72 & 0.98 & 1.00 & 0.98 & 1.00 \\ 
 & B & & High  & 0.86 & 1.03 & 1.06 & 1.02 & 1.00\\ 
 & C & High & Low   & 0.80 & 1.04 & 0.99 & 1.00 & 1.00 \\ 
 & D & & High & 0.88 & 1.00 & 0.95 & 0.97 & 1.00 \\ \hline
 2O2P & A &  Moderate & Low &  0.64 & 0.98 & 0.89 & - & 1.00  \\ 
 & B & & High & 0.79 & 1.00 & 0.92 & - & 1.00

  \\ \hline
\end{tabular}
\end{table}

\begin{table}[ht]
\caption{\label{TabCoverage} Coverage probabilities of estimated 95\% confidence intervals for the simulation study for the two outcomes (2O) and two predictors (2P) scenarios.}
\centering
\begin{tabular}{|l|l|l|l|l|l|l|l|l|l|}
\hline
\textbf{\shortstack{Scen-\\ario}} & 
\textbf{\shortstack{Corr \\ level}} & 
\textbf{\shortstack{Error \\ level}} &  
\textbf{\shortstack{Estim-\\ator}} &  
 & 
\textbf{\shortstack{Strat 1: \\ Case-\\control}} & 
\textbf{\shortstack{Strat 2: \\ Simult-\\aneous}} & 
\textbf{\shortstack{Strat 3: \\ Sequential \\ order 1}} & 
\textbf{\shortstack{Strat 4: \\ Sequential \\ order 2}} & 
\textbf{\shortstack{Strat 5: \\ A-optimal}} \\ 
\hline
 & Low & Low & IPW & $\hat{\beta}_{11}$ &   0.94 & 0.95 & 0.95 & 0.94 & 0.95 \\ 
 & & (A)  &  & $\hat{\beta}_{12}$  & 0.94 & 0.94 & 0.94 & 0.94 & 0.94 \\ \cline{4-10}
 & &  & GR & $\hat{\beta}_{11}$   & 0.95 & 0.94 & 0.95 & 0.94& 0.95\\ 
 & &  & & $\hat{\beta}_{12}$  & 0.95 & 0.93 & 0.94 & 0.93 & 0.93 \\ \cline{3-10}
 & & High & IPW  & $\hat{\beta}_{11}$ &   0.96 & 0.94 & 0.94 & 0.94 & 0.94 \\ 
 & & (B) &  & $\hat{\beta}_{12}$ &   0.95 & 0.95 & 0.94 & 0.95 & 0.95\\ \cline{4-10}
 & &  & GR & $\hat{\beta}_{11}$   & 0.96 & 0.95 & 0.95 & 0.93 & 0.95 \\ 
 2O & &  & & $\hat{\beta}_{12}$   & 0.95 & 0.94 & 0.93 & 0.94 & 0.94 \\ \cline{2-10}
& High & Low & IPW & $\hat{\beta}_{11}$ &   0.94 & 0.94 & 0.95 & 0.94 & 0.95 \\ 
 & & (C) &  & $\hat{\beta}_{12}$  & 0.96 & 0.94 & 0.94 & 0.94 & 0.95 \\ \cline{4-10}
 & &  & GR & $\hat{\beta}_{11}$   & 0.95 & 0.94 & 0.96 & 0.94 & 0.95\\ 
 & &  & & $\hat{\beta}_{12}$  & 0.95 & 0.93 & 0.94 & 0.93 & 0.94 \\ \cline{3-10}
 & & High & IPW  & $\hat{\beta}_{11}$ &   0.95 & 0.94 & 0.95 & 0.93 & 0.94\\ 
 & & (D) &  & $\hat{\beta}_{12}$ &  0.96 & 0.93 & 0.94 & 0.94 & 0.93 \\ \cline{4-10}
 & &  & GR & $\hat{\beta}_{11}$   & 0.95 & 0.94 & 0.95 & 0.93 & 0.92 \\ 
 & &  & & $\hat{\beta}_{12}$   & 0.95 & 0.93 & 0.94 & 0.93 & 0.94 \\ \cline{1-10} 
  & Low & Low & IPW & $\hat{\beta}_{11}$ &  0.95 & 0.94 &0.94 & 0.94 & 0.94 \\ 
 & & (A)  &  & $\hat{\beta}_{12}$  & 0.94 & 0.94 & 0.94 & 0.95 & 0.94 \\ \cline{4-10}
 & &  & GR & $\hat{\beta}_{11}$   & 0.95 & 0.94 & 0.94 & 0.94 & 0.95\\ 
 & &  & & $\hat{\beta}_{12}$  & 0.94 & 0.94 & 0.93 & 0.95 & 0.94 \\ \cline{3-10}
 & & High & IPW  & $\hat{\beta}_{11}$ &  0.94 & 0.94 & 0.95 & 0.94 & 0.94 \\ 
 & & (B) &  & $\hat{\beta}_{12}$ &   0.94 & 0.94 & 0.92 & 0.94 & 0.93\\ \cline{4-10}
 & &  & GR & $\hat{\beta}_{11}$   & 0.95 & 0.94 & 0.94 & 0.93
& 0.94 \\ 
 2P & &  & & $\hat{\beta}_{12}$   & 0.94 & 0.93 & 0.92 & 0.94 &0.93 \\ \cline{2-10}
& High & Low & IPW & $\hat{\beta}_{11}$ &   0.94 & 0.95 & 0.94 & 0.94 & 0.94 \\ 
 & & (C) &  & $\hat{\beta}_{12}$  & 0.94 & 0.94 & 0.93 & 0.94 & 0.94 \\ \cline{4-10}
 & &  & GR & $\hat{\beta}_{11}$   & 0.95 & 0.94 & 0.95 & 0.93 & 0.95 \\ 
 & &  & & $\hat{\beta}_{12}$  & 0.95 & 0.94 & 0.93 & 0.95 & 0.94 \\ \cline{3-10}
 & & High & IPW  & $\hat{\beta}_{11}$ &   0.96 & 0.94 & 0.94 & 0.94 & 0.95\\ 
 & & (D) &  & $\hat{\beta}_{12}$ &  0.95 & 0.93 & 0.93 & 0.94 & 0.94 \\ \cline{4-10}
 & &  & GR & $\hat{\beta}_{11}$   & 0.95 & 0.94 & 0.95 & 0.94 & 0.94 \\ 
 & &  & & $\hat{\beta}_{12}$   & 0.95 & 0.94 & 0.93 & 0.94 & 0.95 \\  \cline{1-10}
\end{tabular}
\centering
\end{table}

\begin{table}[ht]
\caption{\label{TabCoverage2} Coverage probabilities of estimated 95\% confidence intervals for the simulation study for the two outcomes and two predictors (2O2P) scenario.}
\centering
\begin{tabular}{|l|l|l|l|l|l|l|l|l|l|}
\hline
\textbf{\shortstack{Scen-\\ario}} & 
\textbf{\shortstack{Corr \\ level}} & 
\textbf{\shortstack{Error \\ level}} &  
\textbf{\shortstack{Estim-\\ator}} &  
 & 
\textbf{\shortstack{Strat 1: \\ Case-\\control}} & 
\textbf{\shortstack{Strat 2: \\ Simult-\\aneous}} & 
\textbf{\shortstack{Strat 3: \\ Sequential  }} & 
\textbf{\shortstack{Strat 5: \\ A-optimal}} \\ 
\hline
   & & Low & IPW & $\hat{\beta}_{11}$ &   0.96 & 0.95 & 0.95 & 0.95 \\ 
 & & (A)  &  & $\hat{\beta}_{12}$  & 0.94 & 0.94 & 0.94 & 0.95 \\ 
 & &   &  & $\hat{\beta}_{21}$  & 0.95 & 0.95 & 0.94 & 0.94  \\
 & &   &  & $\hat{\beta}_{22}$  & 0.95 & 0.94 & 0.91 & 0.93  \\ \cline{4-9}
 & &  & GR & $\hat{\beta}_{11}$   & 0.95 & 0.94 & 0.94 & 0.95 \\ 
 & &  & & $\hat{\beta}_{12}$  & 0.95 & 0.95 & 0.94 & 0.95 \\
  & &  & & $\hat{\beta}_{21}$  & 0.95 & 0.95 & 0.94 & 0.95  \\
  & &  & & $\hat{\beta}_{22}$  & 0.95 & 0.94 & 0.93 & 0.93 \\ \cline{3-9}
  2O2P& Moderate & High & IPW & $\hat{\beta}_{11}$ &   0.94 & 0.94 & 0.95 & 0.94  \\ 
 & & (B)  &  & $\hat{\beta}_{12}$  & 0.94 & 0.94 & 0.93 & 0.94 \\ 
 & &   &  & $\hat{\beta}_{21}$  & 0.95 & 0.95 & 0.94 & 0.94  \\
 & &   &  & $\hat{\beta}_{22}$  & 0.94 & 0.94 & 0.90 & 0.93  \\ \cline{4-9}
 & &  & GR & $\hat{\beta}_{11}$   & 0.95 & 0.94 & 0.95 & 0.94 \\ 
  & &  & & $\hat{\beta}_{12}$  & 0.95 & 0.94 & 0.94 & 0.94  \\
  & &  & & $\hat{\beta}_{21}$  & 0.95 & 0.94 & 0.95 & 0.94 \\
  & &  & & $\hat{\beta}_{22}$  & 0.95 & 0.94 & 0.92 & 0.94 \\ \cline{3-9}
\cline{1-9}
\end{tabular}
\centering
\end{table}

\begin{table}
\caption{\label{TabScenario2On500} Empirical variance ($\times 10^3$) for each strategy under Scenario 2O: Two outcomes of interest in the supplemental scenario where the phase 2 size is $n = 500$.}
\centering
\begin{tabular}{|l|l|l|l|l|l|l|l|l|}
\hline
\textbf{\shortstack{Corr \\ level}} & 
\textbf{\shortstack{Error \\ level}} &  
\textbf{\shortstack{Estim-\\ator}} &  
 & 
\textbf{\shortstack{Strat 1: \\ Case-\\control}} & 
\textbf{\shortstack{Strat 2: \\ Simult-\\aneous}} & 
\textbf{\shortstack{Strat 3: \\ $\beta_{12}$ last}} & 
\textbf{\shortstack{Strat 4: \\ $\beta_{11}$ last}} & 
\textbf{\shortstack{Strat 5: \\ A-\\optimal}} \\ 
\hline
Low & Low & IPW & var($\hat{\beta}_{11}$) &   11.50 & 5.26 & 5.99 & 4.93 & 5.03\\ 
 & (A) &  & var($\hat{\beta}_{12}$)  & 11.54 & 5.25 & 4.98 & 6.22 & 5.20\\ \cline{3-9}
 &  & GR & var($\hat{\beta}_{11}$)   & 4.02 & 3.32 & 3.89 & 3.14 & 3.45\\
 &  &  & var($\hat{\beta}_{12}$)   & 5.02 & 3.80 & 3.98 & 4.49 & 3.74\\ \cline{2-9}
 & High & IPW  & var($\hat{\beta}_{11}$) &   12.34 & 10.02 & 11.17 & 10.86 & 10.09\\ 
 & (B) &  & var($\hat{\beta}_{12}$) &   11.34 & 8.08 & 8.55 & 9.61 & 8.12\\ \cline{3-9}
 &  & GR & var($\hat{\beta}_{11}$)   &  9.16 & 9.54 & 10.66 & 9.60 & 9.04\\ 
 &  &  & var($\hat{\beta}_{12}$)   & 8.18 & 7.92 & 7.69 & 8.50 & 7.87\\ \hline
High & Low & IPW & var($\hat{\beta}_{11}$) & 15.26 & 8.62 & 9.25 & 7.81 & 8.66\\ 
 & (C)  &  & var($\hat{\beta}_{12}$)  & 15.60 & 7.43 & 6.33 & 8.63 & 6.90\\ \cline{3-9}
 &  & GR & var($\hat{\beta}_{11}$)   & 6.19 & 5.60 & 5.89 & 5.28 & 5.43\\
 &  &  & var($\hat{\beta}_{12}$)   & 6.77 & 5.31 & 4.66 & 6.07 & 5.01\\ \cline{2-9}
 & High & IPW  & var($\hat{\beta}_{11}$) &   16.97 & 16.53 & 17.20 & 16.35 & 15.30\\ 
 & (D) &  & var($\hat{\beta}_{12}$) &  14.14 & 9.62 & 9.49 & 11.82 & 10.30\\ \cline{3-9}
 &  & GR & var($\hat{\beta}_{11}$)   & 13.30 & 15.03 & 15.85 & 15.22 & 14.87\\
 &  &  & var($\hat{\beta}_{12}$)   & 10.91 & 9.44 & 9.19 & 10.74 & 9.32\\ \hline
\end{tabular}
\end{table}

\begin{table}
\caption{\label{TabScenario2Pn500} Empirical variance ($\times 10^3$) for each strategy under Scenario 2P: Two predictors of interest in the supplemental scenario where the phase 2 size is $n = 500$.}
\centering
\begin{tabular}{|l|l|l|l|l|l|l|l|l|l|}
\hline
\textbf{\shortstack{Corr \\ level}} & 
\textbf{\shortstack{Error \\ level}} &  
\textbf{\shortstack{Estim-\\ ator}} &  
 & 
\textbf{\shortstack{Strat 1:\\ Case-\\control}} & 
\textbf{\shortstack{Strat 2:\\ Simult-\\aneous}} & 
\textbf{\shortstack{Strat 3: \\ $\beta_{22}$ last}} & 
\textbf{\shortstack{Strat 4: \\ $\beta_{12}$ last}} & 
\textbf{\shortstack{Strat 5: \\ A-\\optimal}} \\ 
\hline
Low& Low & IPW & var($\hat{\beta}_{12}$) &   6.38 & 4.67 & 4.83 & 4.67 & 4.65 \\ 
 & (A) &  & var($\hat{\beta}_{22}$)  & 8.16 & 6.04 & 5.69 & 5.97 & 5.84 \\ \cline{3-9}
 &  & GR & var($\hat{\beta}_{12}$)   & 3.93 & 3.64 & 3.60 & 3.58 & 3.38 \\ 
 &  & & var($\hat{\beta}_{22}$)   & 5.81 & 5.09 & 5.16 & 5.15 & 5.05 \\ \cline{2-9}
 & High & IPW  & var($\hat{\beta}_{12}$) &   6.94 & 5.93 & 5.65 & 6.03 & 5.92 \\ 
 & (B) &  & var($\hat{\beta}_{22}$) &  8.43 & 6.87 & 6.82 & 6.90 & 7.23 \\ \cline{3-9}
 &  & GR & var($\hat{\beta}_{12}$)   & 5.57 & 5.34 & 5.28 & 5.33 & 5.21 \\ 
 &  & & var($\hat{\beta}_{22}$)   & 6.98 & 6.51 & 6.51 & 6.54 & 6.67 \\ \hline
High & Low & IPW & var($\hat{\beta}_{12}$) &   11.14 & 8.56 & 8.90 & 9.05 & 9.00 \\ 
 & (C) &  & var($\hat{\beta}_{22}$)  & 12.60 & 9.67 & 10.26 & 9.95 & 10.00 \\ \cline{3-9}
 &  & GR & var($\hat{\beta}_{12}$)   & 7.74 & 7.05 & 7.14 & 7.13 & 6.68\\ 
 &  & & var($\hat{\beta}_{22}$)   & 10.02 & 9.09 & 9.02 & 8.69 & 8.36 \\ \cline{2-9}
 & High & IPW  & var($\hat{\beta}_{12}$) &   11.49 & 10.42 & 10.87 & 10.90 & 10.63 \\ 
 & (D) &  & var($\hat{\beta}_{22}$) &  13.55 & 11.72 & 12.49 & 11.84 & 11.52 \\ \cline{3-9}
 &  & GR & var($\hat{\beta}_{12}$)   & 9.34 & 10.00 & 9.52 & 9.78 & 9.51 \\ 
 &  & & var($\hat{\beta}_{22}$)   & 11.51 & 11.16 & 11.27 & 10.99 & 10.70 \\ \hline
\end{tabular}
\end{table}

\begin{table}[ht]
\caption{\label{TabCoveragen500} Coverage probabilities of estimated 95\% confidence intervals for the simulation study for the two outcomes (2O) and two predictors (2P) scenarios in the supplemental scenario where the phase 2 size is $n = 500$.}
\centering
\begin{tabular}{|l|l|l|l|l|l|l|l|l|l|}
\hline
\textbf{\shortstack{Scen-\\ario}} & 
\textbf{\shortstack{Corr \\ level}} & 
\textbf{\shortstack{Error \\ level}} &  
\textbf{\shortstack{Estim-\\ator}} &  
 & 
\textbf{\shortstack{Strat 1: \\ Case-\\control}} & 
\textbf{\shortstack{Strat 2: \\ Simult-\\aneous}} & 
\textbf{\shortstack{Strat 3: \\ Sequential \\ order 1}} & 
\textbf{\shortstack{Strat 4: \\ Sequential \\ order 2}} & 
\textbf{\shortstack{Strat 5: \\ A-optimal}} \\ 
\hline
 & Low & Low & IPW & $\hat{\beta}_{11}$ &   0.94 & 0.94 & 0.94 & 0.94 & 0.95\\
 & & (A)  &  & $\hat{\beta}_{12}$  & 0.95 & 0.94 & 0.93 & 0.94 & 0.94\\ \cline{4-10}
 & &  & GR & $\hat{\beta}_{11}$   &  0.95 & 0.94 & 0.95 & 0.94 & 0.94\\ 
 & &  & & $\hat{\beta}_{12}$  &  0.95 & 0.93 & 0.91 & 0.93 & 0.93\\ \cline{3-10}
 & & High & IPW  & $\hat{\beta}_{11}$ &   0.95 & 0.94 & 0.94 & 0.90 & 0.93\\ 
 & & (B) &  & $\hat{\beta}_{12}$    & 0.95 & 0.93 & 0.92 & 0.94 & 0.93\\ \cline{4-10}
 & &  & GR & $\hat{\beta}_{11}$   & 0.95 & 0.93 & 0.94 & 0.92 & 0.94\\ 
 2O & &  & & $\hat{\beta}_{12}$   & 0.95 & 0.92 & 0.92 & 0.94 & 0.93\\ \cline{2-10}
& High & Low & IPW & $\hat{\beta}_{11}$ &    0.94 & 0.94 & 0.94 & 0.93 & 0.94\\ 
 & & (C) &  & $\hat{\beta}_{12}$  & 0.95 & 0.93 & 0.95 & 0.94 & 0.94\\ \cline{4-10}
 & &  & GR & $\hat{\beta}_{11}$   & 0.95 & 0.94 & 0.94 & 0.93 & 0.94\\ 
 & &  & & $\hat{\beta}_{12}$  & 0.95 & 0.93 & 0.93 & 0.92 & 0.92\\ \cline{3-10}
 & & High & IPW  & $\hat{\beta}_{11}$ &   0.95 & 0.93 & 0.93 & 0.91 & 0.93\\ 
 & & (D) &  & $\hat{\beta}_{12}$ & 0.95 & 0.94 & 0.94 & 0.94 & 0.94\\ \cline{4-10}
 & &  & GR & $\hat{\beta}_{11}$   & 0.95 & 0.92 & 0.94 & 0.91 & 0.92\\ 
 & &  & & $\hat{\beta}_{12}$   & 0.95 & 0.93 & 0.92 & 0.93 & 0.93\\ \cline{1-10} 
  & Low & Low & IPW & $\hat{\beta}_{11}$ &   0.94 & 0.93 & 0.94 & 0.94 & 0.95 \\ 
 & & (A)  &  & $\hat{\beta}_{12}$  & 0.94 & 0.94 & 0.92 & 0.94 & 0.93\\ \cline{4-10}
 & &  & GR & $\hat{\beta}_{11}$   & 0.95 & 0.94 & 0.93 & 0.94 & 0.94\\
 & &  & & $\hat{\beta}_{12}$  & 0.94 & 0.94 & 0.93 & 0.95 & 0.93\\ \cline{3-10}
 & & High & IPW  & $\hat{\beta}_{11}$ &  0.95 & 0.94 & 0.94 & 0.93 & 0.93\\ 
 & & (B) &  & $\hat{\beta}_{12}$ &   0.95 & 0.93 & 0.92 & 0.93 & 0.93\\ \cline{4-10}
 & &  & GR & $\hat{\beta}_{11}$   &  0.95 & 0.94 & 0.95 & 0.93 & 0.93\\ 
 2P & &  & & $\hat{\beta}_{12}$   & 0.95 & 0.93 & 0.92 & 0.93 & 0.93\\ \cline{2-10}
& High & Low & IPW & $\hat{\beta}_{11}$ &   0.95 & 0.94 & 0.94 & 0.93 & 0.94\\ 
 & & (C) &  & $\hat{\beta}_{12}$  &0.95 & 0.93 & 0.93 & 0.95 & 0.94\\ \cline{4-10}
 & &  & GR & $\hat{\beta}_{11}$   &  0.95 & 0.94 & 0.93 & 0.94 & 0.95\\ 
 & &  & & $\hat{\beta}_{12}$  & 0.95 & 0.94 & 0.92 & 0.94 & 0.94\\ \cline{3-10}
 & & High & IPW  & $\hat{\beta}_{11}$ &  0.95 & 0.93 & 0.93 & 0.93 & 0.93\\ 
 & & (D) &  & $\hat{\beta}_{12}$ &  0.94 & 0.93 & 0.92 & 0.94 & 0.92\\ \cline{4-10}
 & &  & GR & $\hat{\beta}_{11}$   & 0.95 & 0.94 & 0.93 & 0.93 & 0.94\\ 
 & &  & & $\hat{\beta}_{12}$   & 0.94 & 0.93 & 0.92 & 0.93 & 0.93\\  \cline{1-10}
\end{tabular}
\centering
\end{table}

\renewcommand{\arraystretch}{1}

\newpage


\label{lastpage}
\end{document}